\documentclass[10pt,conference,letterpaper]{IEEEtran}
\usepackage{times,amsmath,amsthm,epsfig}

\usepackage{graphicx}
\usepackage{balance} 
\usepackage{xcolor}
\usepackage{subcaption}
\usepackage{mdframed}
\usepackage{latexsym}
\usepackage{amssymb}
\usepackage{tikz}
\usetikzlibrary{shapes,shadows,arrows,calc,patterns,positioning}
\usepackage{pgfplots}

\newcommand{\adapt}{\mbox{\textsc{Cobra}}}

\definecolor{darkblue}{rgb}{0.0, 0.0, 0.55}
\definecolor{darkmidnightblue}{rgb}{0.0, 0.0, 0.4}
\definecolor{cobalt}{rgb}{0.0, 0.28, 0.67}
\definecolor{dukeblue}{rgb}{0.0, 0.0, 0.8}
\definecolor{myDarkGreen}{rgb}{0.0, 0.42, 0.24}
\definecolor{darkgreen}{rgb}{0.0, 0.8, 0.0}

\newcommand{\review}[1]{\textcolor{black}{#1}}
\newcommand{\eat}[1]{}

\newcommand{\kw}[1]{{\textbf{#1}}} %keyword
\newcommand{\cmnt}[1]{\emph{#1}} %comment
\newcommand{\lnum}[1]{\textcolor{gray}{{\footnotesize{#1}}}} %line number
\newcommand{\str}[1]{#1} %string in program
\newcommand{\myAnd}{{\small{AND}}}
\newcommand{\myOr}{{\small{OR}}}
\newcommand{\andOrDag}{{\small{AND-OR DAG}}}
\newcommand{\myDag}{{\small{DAG}}}
\newcommand{\myFir}{\mbox{\small{F-IR}}}
\newcommand{\regionDag}{Region \myDag{}}

\newcommand{\shiftLeftTiny}{-0.3em}
\newcommand{\shiftRightTiny}{0.3em}
\newcommand{\shiftLeftSmallest}{-0.6em}
\newcommand{\shiftRightSmallest}{0.6em}

\newcommand{\shiftLeftSmaller}{-1.25em}
\newcommand{\shiftRightSmaller}{1.25em}
\newcommand{\shiftLeftSmall}{-2.5em}
\newcommand{\shiftRightSmall}{2.5em}
\newcommand{\shiftLeftSmallPlus}{-3.5em}

\newcommand{\shiftLeft}{-5em}
\newcommand{\shiftRight}{5em}
\newcommand{\shiftLeftBig}{-7.5em}
\newcommand{\shiftRightBig}{7.5em}

\newcommand{\shiftRightBigger}{10em}
\newcommand{\shiftLeftBiggest}{-12.5em}

\tikzstyle{orNode} = [draw,rectangle, minimum height=5mm, minimum width=5mm]
\tikzstyle{andNode} = [draw,circle]
\tikzstyle{line} = [draw]
\tikzstyle{noBorderNode} = [rectangle, minimum height=5mm]
\tikzstyle{orNodeHl} = [draw=blue,rectangle, minimum height=5mm, minimum width=5mm]
\tikzstyle{andNodeHl} = [draw=blue,circle]
\tikzstyle{lineHl} = [draw=blue]
\newcommand{\regionWidth}{0.25em}

\newcommand{\bls}{\baselineskip}
\tikzstyle{basicBlock} = [draw, minimum width=\regionWidth]
\tikzstyle{loopRegion} = [draw,pattern=horizontal lines, minimum width=\regionWidth, pattern color=blue]
\tikzstyle{seqRegion} = [draw,pattern=crosshatch, minimum width=\regionWidth, pattern color=myDarkGreen]
\tikzstyle{condRegion} = [draw, pattern=dots, minimum width=\regionWidth]
\tikzstyle{funcRegion} = [draw, pattern=north east lines, minimum width=\regionWidth]
\tikzstyle{myStage} = [draw, rectangle, rounded corners, text width=7em, text centered]
\tikzstyle{myProcess} = []
\newcommand{\mli}[1]{\mathit{#1}}

\title{\adapt{}: A Framework for\\Cost Based Rewriting of Database Applications (Extended Report)}

\author{%
    {K. Venkatesh Emani, S. Sudarshan}
    \vspace{1.6mm}\\
    \fontsize{10}{10}\selectfont\itshape
    IIT Bombay, India\\
    \fontsize{9}{9}\selectfont\ttfamily\upshape
    venkateshek@cse.iitb.ac.in, sudarsha@cse.iitb.ac.in%
}

\begin{document}
\maketitle

\begin{abstract}
Database applications are typically written using a mixture of imperative languages and declarative frameworks for data processing. Application logic gets distributed across the declarative and imperative parts of a program. Often, there is more than one way to implement the same program, whose efficiency may depend on a number of parameters. In this paper, we propose a framework that automatically generates all equivalent alternatives of a given program using a given set of program transformations, and chooses the least cost alternative. We use the concept of program regions as an algebraic abstraction of a program and extend the Volcano/Cascades framework for optimization of algebraic expressions, to optimize programs. We illustrate the use of our framework for optimizing database applications. We show through experimental results, that our framework has wide applicability in real world applications and provides significant performance benefits.
\end{abstract}
\section{Introduction}
\label{sec:intro}
Database applications are typically written using a mixture of imperative languages such as Java for business logic, and declarative frameworks for data processing. Examples of such frameworks include SQL (JDBC) with Java, object-relational mappers (ORMs), large scale data processing frameworks such as Apache Spark, and Python data science libraries (example: pandas), among others. These frameworks provide high level operators/library functions for expressing common data processing operations, and contain efficient implementations of these functions. 

However, in many applications, data processing operations are often (partially) implemented in imperative code. The reasons for this include modularity,  limited framework expertise of the developer, need for custom operations that cannot be expressed in the declarative framework, etc. Consequently, data processing is distributed across the imperative and declarative parts of the application. Often, there is more than one way to implement the same program, and the best approach may be chosen depending on a number of parameters.

\begin{figure}
	\centering
	\includegraphics[width=0.35\textwidth]{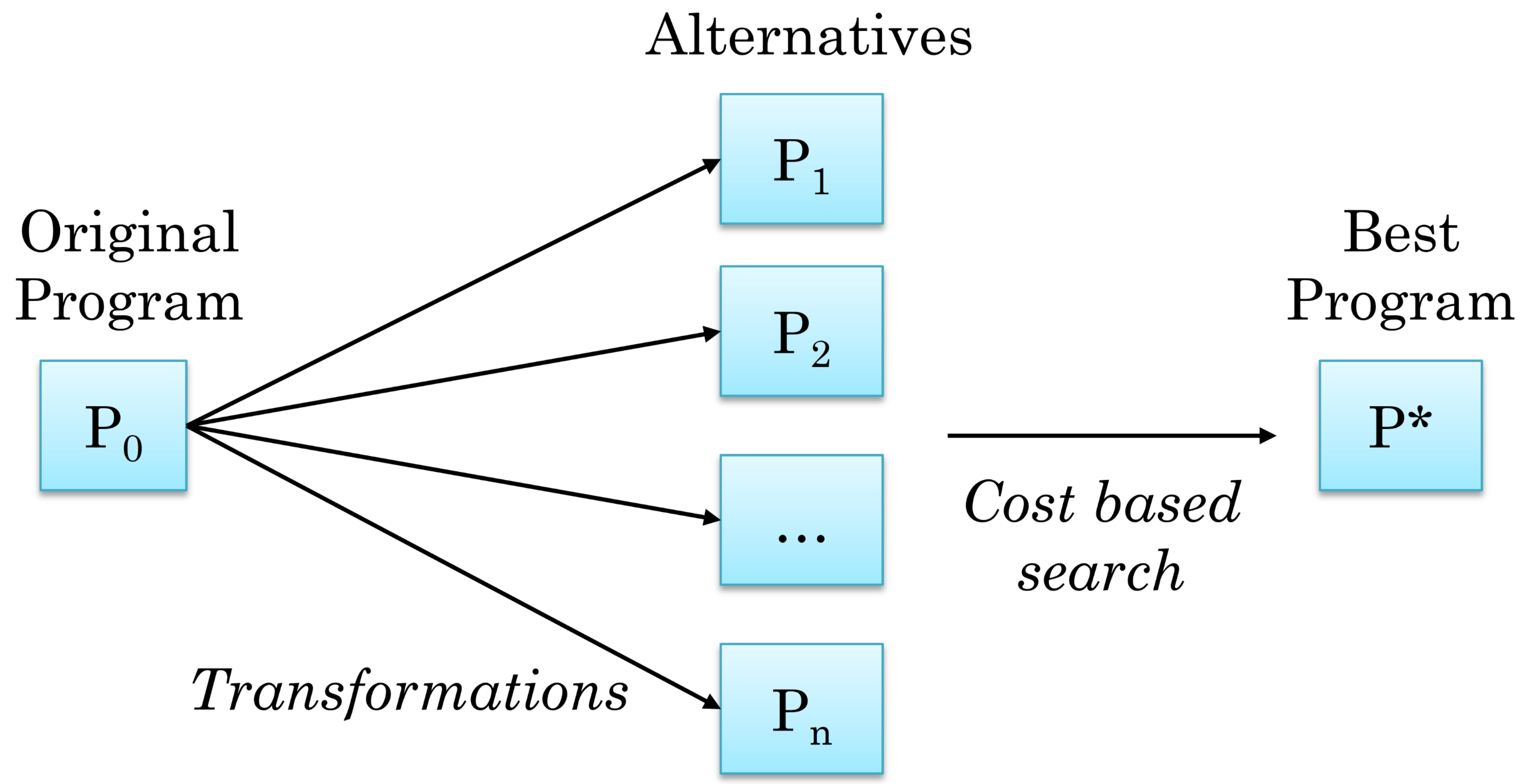}
	\caption{\adapt{} Illustration}
	\label{fig:overview}
\end{figure}

This raises an interesting question for an optimizing compiler for data processing applications. Given an application program, is it possible to generate semantically equivalent alternatives of the program using program transformations, and choose the program with the least cost depending on the context? In this paper we propose the \adapt{}\footnote{
	Acronym formed from COst Based Rewriting of (database) Applications.}
framework to achieve this, as illustrated in Figure~\ref{fig:overview}.

There has been work on rewriting data processing programs for improved performance using program transformations~\cite{Cheung13, CHA11, TKDE14, EMANI16, Radoi14}. However, existing techniques fail to consider all possible alternatives for cost based rewriting. They either apply the proposed transformations in a specific order~\cite{CHA11}, or carefully craft the transformation rules so that the rule set is confluent and terminating~\cite{EMANI16}. This is not a viable solution for all rule sets, especially as the number/complexity of rules increases. A brute force solution is to keep applying all possible transformations as long as any one of them is applicable; however, this may cause the transformation process to never terminate, in case of cyclic transformation rules. For example, in their work on translating imperative code to map-reduce~\cite{Radoi14}, Radoi et al. state that their transformation rules are neither confluent nor terminating, and use a heuristic driven by a cost function to guide the search for possible rewrites. However, such an approach in general has the disadvantage of missing out on useful rewrites that are not considered by the heuristic.

A similar problem has been solved for the purpose of query optimization in databases. Graefe et al. proposed the Vol\-cano/Cas\-cades framework~\cite{Volcano,Goetz95}, which uses an \andOrDag{} representation (details in Section~\ref{sec:background}) to enumerate all alternative rewrites for a given SQL query (relational algebra expression) generated using transformation rules, and to choose the best query (plan) by searching through the space of possible rewrites. Although designed for query optimization, the Vol\-cano/Cas\-cades framework can be used with any algebra.

Such a framework can be used with program transformations based on expressions, as described in~\cite{peggy}. Examples of such transformations include many peephole optimizations such as constant folding, strength reduction, etc. However, transformations proposed for optimizing data processing applications typically involve rewriting conditional statements, loops, functions, or even the entire program. Such transformations involving larger program units are not amenable for direct integration into an algebraic framework like Vol\-cano/Cas\-cades.

In this paper, we identify that \emph{program regions}~\cite{Muchnick98}, which we used for transformations in our previous work~\cite{EMANI16}, provide a natural abstraction for dividing an imperative program into parts, which can then be optimized individually and collectively using an extension of the Volcano/Cas\-cades framework. Program regions are structured fragments of a program such as straight line code, if-else, loops, functions, etc. (details in Section~\ref{sec:background}). Our framework, \adapt{}, represents a program as an \andOrDag{} using program regions. Program transformations add alternatives to this \andOrDag{}. \adapt{} can be used for cost-based transformations in any program with well-defined program regions. However, in this paper, we restrict our attention to the use of \adapt{} for optimizing database applications.

Our contributions in this paper are as follows:
\begin{itemize}
    \item We describe the \andOrDag{} representation of an imperative program with regions, and discuss how the alternatives generated using program transformations are represented using the \andOrDag{} (Section~\ref{sec:and_or_rep}).
    \item We illustrate the use of our framework for optimizing database applications. To this end, we discuss an intermediate representation for transformations in database applications (Section~\ref{sec:ir}) building on our earlier work~\cite{EMANI16}.
	\item We present a cost model (Section~\ref{sec:costing}) to estimate the cost of database application programs, with a focus on cost of query execution statements and loops over query results.
    \item We built the \adapt{} optimizer by incorporating our techniques into a system that implements the Volcano/Cascades framework. We present an experimental evaluation (Section~\ref{sec:experiments}) of \adapt{} on a real world application, to show the applicability of our techniques and their impact on application performance.
\end{itemize}

We present a motivating example in Section~\ref{sec:motivation}, and discuss the necessary background in Section~\ref{sec:background}. We discuss related work in Section~\ref{sec:relwork}, and conclude in Section~\ref{sec:conclusion}.

\section{Motivating Example}
\label{sec:motivation}
The \adapt{} framework can be used for optimizing programs using a variety of data access methods such as JDBC, web services, object relational mappers (ORM) etc. In this section we discuss an example program that uses the Hibernate ORM~\cite{HIB}, to motivate the need for \adapt{}. 

Object relational mapping frameworks such as Hibernate enable access to the database using the same language as the application~\cite{Cheung13} without writing explicit SQL queries. The framework automatically generates relevant queries from object accesses and translates query results into objects, based on a specified mapping between database tables and application classes. 

\begin{figure}
    \small
    \input{snippets/ormMapping}
    \caption{Hibernate object-relation mapping specification}
    \label{fig:hib_schema_mapping}
\end{figure}
For example consider Figure~\ref{fig:hib_schema_mapping}, which shows a schema definition in the Hibernate ORM.
The class \emph{Order} is mapped to the database table \emph{orders}. When \emph{Order} objects are retrieved, the framework implicitly creates a query on \emph{orders}, and populates the attributes of \emph{Order}. The relationship from table \emph{orders} to table \emph{customers} (mapped by class \emph{Customer}) is expressed as an attribute of \emph{Order}.

Objects (rows) retrieved from the database are cached upon first access using their id (primary key). Thereafter, these objects can be accessed inside the application without having to query the database again. Hibernate supports lazy loading, i.e., fetching an attribute of an object only when the attribute is accessed; this facilitates fetching information from a related table (such as \emph{customer} in \emph{Order}) only when needed. Most ORMs also allow users to express complex queries using SQL or object based query languages.

ORMs are widely used in OLTP applications~\cite{Cheung13}, and their use in reporting applications is not uncommon~\cite{JasperHibernate}. Inefficiencies due to the usage of ORM frameworks are also well known~\cite{Chen2016}, and have been addressed by earlier optimization techniques~\cite{Cheung13, EMANI16} (refer related work, Section~\ref{sec:relwork}).

\begin{figure}
    \begin{subfigure}{0.45\textwidth}
        \small
        \texttt{\input{snippets/hib}}
        \vspace*{-4pt}
        \caption{$P_0$: Program using Hibernate ORM}
        \label{fig:hib}
    \end{subfigure}
    \hrule

    \begin{subfigure}{0.45\textwidth}
        \vspace*{0.5em}
        \small
        \texttt{\input{snippets/join_db}}
        \vspace*{-4pt}
        \caption{$P_1$: $P_0$ rewritten to use Hibernate SQL query API}
        \label{fig:join_db}
    \end{subfigure}
    \hrule

    \begin{subfigure}{0.45\textwidth}
        \vspace*{0.5em}
        \small
        \texttt{\input{snippets/join_app}}
        \vspace*{-4pt}
        \caption{$P_2$: $P_0$ rewritten to use prefetching}
        \label{fig:join_app}
    \end{subfigure}
    \hrule

    \vspace*{0.5em}
    \caption{Alternative implementations of the same program}
    \label{fig:alt}
\end{figure}

Figure~\ref{fig:hib} shows a sample program using the Hibernate ORM that processes a list of orders, along with customer related information. The program uses an ORM API (\textit{loadAll}) to fetch all \emph{Orders} objects, and then processes each order inside a loop. However, for each order, the framework generates a separate query to fetch the related \emph{customer} information, which resides in another table. This causes a lot of network round trips, leading to poor performance. This issue is known as the \emph{N+1 select problem} in ORMs~\cite{Chen2016}.

To avoid this problem, a join query is usually suggested to fetch the required data, while restricting the number of queries to one. This is shown in program $P_1$ in Figure~\ref{fig:join_db}%
\footnote{%
We use a pseudo function \emph{executeQuery} that takes a query, executes it and returns the results as a collection of objects. Also, variable types have not been dis\-played for ease of presentation. Our implementation uses the actual source code.}.
$P_1$ follows the general rule of thumb where  data processing is pushed into the database as much as possible, thus allowing the database to use clever execution plans to minimize query execution time.

The join query shown in $P_1$ may lead to duplication of the customers rows in the join result (as each customer typically places multiple orders). For small data sizes or a few rows when the \emph{orders} fetched are filtered using a selection, this duplication may not have a significant impact. However, for higher cardinalities, the join result may be large and transferring the results over a slow remote network from the database to the application may incur significant latency.
In such cases, an equivalent program $P_2$ shown in Figure~\ref{fig:join_app}%
\footnote{\label{footnote:cache}The pseudo function \emph{cacheByColumn} caches a query result collection based on the value of a given column as key, and \emph{lookup\-Cache} fetches a value from the cache using a given key.
Cache may be in the form a simple hashmap or use caching frameworks such as Memcache or EhCache, which are used by many applications for client side query result caching.
ORM frameworks such as Hibernate provide caching implicitly.}
may be faster, provided the tables \emph{orders} and \emph{customers} fit in the application server memory.
This is because $P_2$ fetches individual tables and performs a join at the application, thus avoiding transfer of a large amount of data over the network.

Existing approaches for rewriting ORM applications with SQL, such as ~\cite{EMANI16, Cheung13} apply transformations with the sole aim of pushing data processing to the database; thus, they transform $P_0$ to $P_1$. Other transformations, such as prefetching query results~\cite{RAM12} may be used to transform $P_0$ to $P_2$. However, neither $P_1$ nor $P_2$ is the best choice in all situations. Using \adapt{}, all alternatives such as $P_1$, $P_2$, and others can be generated using program transformations proposed earlier~\cite{Cheung13, EMANI16, RAM12, TKDE14}, and the best program can be chosen in a cost-based manner.
\section{Background}
\label{sec:background}
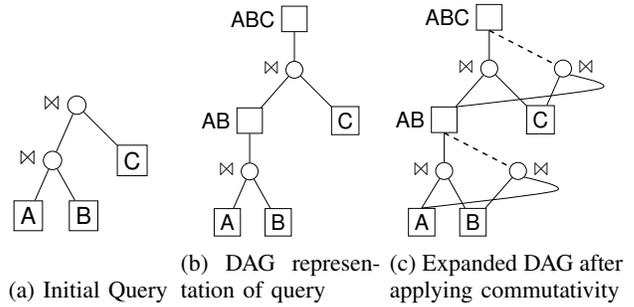
\begin{figure}
\centering
\begin{subfigure}[b]{0.25\linewidth}
       \centering
       \resizebox{0.85\textwidth}{!}{\begin{tikzpicture}[font=\sffamily\large]

\node [andNode, label=west:$\Join$] 
(abc)
{};

\node [andNode, below of=abc, xshift=\shiftLeftSmaller, label=west:$\Join$] 
(ab)
{};
\node [orNode, right of=ab, xshift=\shiftRightSmaller] 
(c)
{C};

\node [orNode, below of=ab, xshift=\shiftLeftSmaller] 
(a)
{A};
\node [orNode, right of=a] 
(b)
{B};

\path (abc) edge (ab);
\path (abc) edge (c);
\path (ab) edge (a);
\path (ab) edge (b);
\end{tikzpicture}}
       \vspace*{12pt}
       \caption{Initial Query}
       \label{fig:join_abc_tree}
\end{subfigure}
\begin{subfigure}[b]{0.3\linewidth}
       \centering
       \resizebox{0.85\textwidth}{!}{\begin{tikzpicture}[font=\sffamily\large]

\node [orNode, label=west:ABC] 
(abc)
{};

\node [andNode, below of=abc, label=west:$\Join$] 
(abcOp)
{};

\node [orNode, below of=abcOp, xshift=\shiftLeftSmall, label=west:{AB}] 
(ab)
{};
\node [orNode, right of=ab, xshift=\shiftRightSmall] 
(c)
{C};

\node [andNode, below of=ab, label = west:$\Join$] 
(abOp)
{};

\node [orNode, below of=abOp, xshift=\shiftLeftSmaller] 
(a)
{A};
\node [orNode, right of=a] 
(b)
{B};

\path (abc) edge (abcOp);
\path (abcOp) edge (ab);
\path (abcOp) edge (c);
\path (ab) edge (abOp);
\path (abOp) edge (a);
\path (abOp) edge (b);

\end{tikzpicture}}
       \caption{DAG represen\-tation of query}
       \label{fig:query_and_or_initial}
\end{subfigure}
\begin{subfigure}[b]{0.35\linewidth}
       \centering
       \resizebox{\textwidth}{!}{\begin{tikzpicture}[font=\sffamily\large]

\node [orNode, label=west:ABC] 
(abc)
{};

\node [andNode, below of=abc, label=west:$\Join$] 
(abcOp)
{};
\node [andNode, right of=abcOp, xshift=\shiftRightSmaller, label=0:$\Join$] 
(abcOp2)
{};

\node [orNode, below of=abcOp, xshift=\shiftLeftSmall, label=west:{AB}] 
(ab)
{};
\node [orNode, right of=ab, xshift=\shiftRightSmall] 
(c)
{C};

\node [andNode, below of=ab, label = west:$\Join$] 
(abOp)
{};
\node [andNode, right of=abOp, xshift=\shiftRightSmaller, label=0:$\Join$] 
(abOp2)
{};

\node [orNode, below of=abOp, xshift=\shiftLeftSmaller] 
(a)
{A};
\node [orNode, right of=a] 
(b)
{B};

\path (abc.south) edge (abcOp);
\path [dashed, thick] (abc.south) edge (abcOp2);
\path (abcOp) edge (ab);
\path (abcOp) edge (c);
\path (ab) edge (abOp);
\path (abOp) edge (a.north);
\path (abOp) edge (b);
\path(abcOp2) edge (c);
\path [dashed, thick] (ab.south) edge (abOp2);
\path (abOp2) edge (b);

\draw plot [smooth] coordinates  { (abcOp2.south east) ($ (abcOp2.south east) + (1.75em,-0.75em) $) (ab.north east)};
\draw plot [smooth] coordinates  { (abOp2.south east) ($ (abOp2.south east) + (1.75em,-0.75em) $) (a.north)};

\end{tikzpicture}}
       \caption{Expanded DAG after applying commutativity}
       \label{fig:query_and_or_expanded}
\end{subfigure}
\caption{Representing alternative query rewrites using the \andOrDag{}}
\label{fig:join_abc_and_or_all}
\end{figure}

In this section, we give a background of (a) the \andOrDag{} representation for cost based query optimization in the Vol\-cano/Cas\-cades framework, and (b) program regions.

\subsection{Volcano/Cas\-cades AND-OR DAG}
\label{sec:volcano_and_or}
Our discussion of \andOrDag{}s is based on~\cite{ROY00}. An \andOrDag{} is a directed acyclic graph where each node in the graph is classified as one of two types: an \myAnd{} node, or an \myOr{} node. The children of an \myOr{}-node can only be \myAnd{}-nodes, and vice versa. In the case of relational algebra expressions (queries), \myAnd{} nodes represent operators, and \myOr{} nodes represent relations. For example, consider the join query $(A \Join B) \Join C$, which is shown as a tree in Figure~\ref{fig:join_abc_tree}. The \andOrDag{} representation for this query is shown in Figure~\ref{fig:query_and_or_initial}.

The Volcano framework for optimization of algebraic expressions is based on equivalence rules. This framework allows the optimizer implementor to specify transformation rules that state the equivalence of two algebraic expressions; examples of such rules include join commutativity ($A \Join B \leftrightarrow B \Join A$) and join associativity ($(A \Join B) \Join C \leftrightarrow A \Join (B \Join C)$), in the case of query optimization. Transformation rules are applied on an expression; while new expressions are added, the old ones are retained in the \andOrDag{}. 

Each \myOr{}-node can have multiple children representing alternative ways of computing the same result, while each \myAnd{}-node represents the root operator of a tree that computes the result. For the query $(A \Join B) \Join C$, the \andOrDag{} after applying commutativity is shown in Figure~\ref{fig:query_and_or_expanded}. The alternatives added are shown using a dotted line connecting the OR node to the root operator of the new expression. Thus, we obtain the following alternatives for the root OR node:  $(A \Join B) \Join C$, $ (B \Join A) \Join C$, $C \Join (A \Join B)$, and $C \Join (B \Join A)$. Note that the commutativity transformation is cyclic. The Volcano/Cas\-cades framework has efficient techniques for identifying duplicates, so the transformation process will terminate even in the presence of cyclic transformations.

Each operator in the DAG may be implemented using one of a few alternatives. For example, a join operator may be implemented using a hash join, indexed nested loops join, or a merge join. This adds further alternatives to the \andOrDag{} (not shown in Figure~\ref{fig:join_abc_and_or_all}).
The cost of any node in the \andOrDag{} is calculated using cost of child nodes, as shown in the table below. 
\begin{center}
\small
\begin{tabular}{|l|l|}
	\hline
	\textbf{Node type} & \textbf{\hspace*{2cm}Cost formula} \\\hline
	\myOr{} node & Minimum of cost of each child\\
	~ & (base case: single relation)\\\hline
	\myAnd{} node & Cost of operator + Sum of costs of children\\\hline
\end{tabular}\\	
\end{center}
The plan corresponding to the least cost at the root node of the \andOrDag{} is the optimized plan. 

In the case of query optimization, the cost assigned to a particular node
depends on factors such as the number of rows in the relation, the type of the operator and its implementation, presence of indexes etc. We skip further details of costing for query optimization and refer the reader to~\cite{Volcano, Goetz95}.

\subsection{Program regions}
\label{subsec:regions}
A region is any structured fragment in a program with a single entry and single exit~\cite{Hall95}. Examples of regions include a single statement (\emph{basic block region}), if-else (\emph{conditional region}), loop (\emph{loop region}), etc. A sequence of two or more regions is called a \emph{sequential region}\footnote{%
    Some approaches consider a basic block region as a sequence of statements. In this paper, we consider each statement as a basic block, and treat a sequence of statements as a sequential region consisting of basic blocks. In our implementation, we use an intermediate representation of bytecode~\cite{soot}, where each statement is represented using a three-address code~\cite{Aho06}.}%
. Regions can contain other regions, so they present a hierarchical view of the program. The contained region is called a \emph{sub-region} and the containing region is called the \emph{parent region}. The outermost region represents the entire program. 

\begin{figure}
    \small
    \begin{subfigure}{0.45\textwidth}
        \begin{minipage}{0.2\textwidth}
            \begin{subfigure}{\textwidth}
                \begin{tikzpicture}
%reference - should be long enough to cover the entire picture
\node[basicBlock,opacity=0,minimum height=9\baselineskip] 
(ref) {};

%basic blocks
\node[basicBlock,minimum height=0.85\bls,left of=ref, yshift=2.8\baselineskip]
(b2) {};

\node[basicBlock,minimum height=0.85\bls,below of=b2, yshift=0.85\baselineskip]
(b3) {};

\node[basicBlock,minimum height=0.85\bls,below of=b3, yshift=1.85\baselineskip]
(b4) {};

\node[basicBlock,minimum height=0.85\bls,below of=b4, yshift=1.85\baselineskip]
(b5) {};

\node[basicBlock,minimum height=0.85\bls,below of=b5, yshift=1.85\baselineskip]
(b6) {};

\node[seqRegion,minimum height=3\bls, left of=b3, xshift=2.2em, yshift=-2\baselineskip]
(s46) {};

\node[loopRegion,minimum height=5.1\bls, left of=b3, xshift=1.3em, yshift=-2\baselineskip]
(l37) {};

\node[seqRegion,minimum height=7.2\baselineskip,left of=ref, xshift=-2.8em, yshift=-0.25\bls]
(s27) {};

%demarking lines
\draw[myDarkGreen] (s46.north west) -- ($ (s46.north) + (7.3, 0) $);
\draw[myDarkGreen] (s46.south west) -- ($ (s46.south) + (7.3, 0) $);
%S1
\draw[blue] (l37.north west) -- ($ (l37.north) + (7.6, 0) $);
\draw[blue] (l37.south west) -- ($ (l37.south) + (7.6, 0) $);
%L1
\draw[myDarkGreen] (s27.north west) -- ($ (s27.north) + (7.95, 0) $);
\draw[myDarkGreen] (s27.south west) -- ($ (s27.south) + (7.95, 0) $);

\end{tikzpicture} 
            \end{subfigure}
        \end{minipage}
        ~
        \hspace*{-3.1em}
        \begin{subfigure}{\textwidth}
            \include{snippets/hib_with_lnums}
        \end{subfigure}
        \hrule

        \begin{subfigure}{\textwidth}
            \include{figures/regions_legend}
        \end{subfigure}
    \end{subfigure}

    \caption{Program regions for program $P_0$ from Figure~\ref{fig:hib}}
    \label{fig:regions}
\end{figure}

For example consider Figure~\ref{fig:regions}, which replicates the program $P_0$ from Figure~\ref{fig:hib} with program regions shown alongside the code (note the naming convention for regions). The outermost region in Figure~\ref{fig:regions} is a sequential region $P_0.S_{2-7}$, which consists of basic block $P_0.B_2$ followed by a loop region $P_0.L_{3-7}$. The loop region in turn is composed of a basic block $P_0.B_3$ and a sequential region $P_0.S_{4-6}$, and so on (breakup of $P_0.S_{4-6}$ into its basic blocks is not shown).

\review{Regions can be built from the control flow graph (CFG) using rules described in~\cite{Muchnick98}. We use this approach in our implementation. Alternatively, it is possible to use an abstract syntax tree of code written in a structured programming language to identify program regions. Exceptions may violate the normal control flow in a region. Currently, our techniques do not preserve exception behavior in the program; handling this is part of future work.}

\section{\andOrDag{} Representation\\of Programs}
\label{sec:and_or_rep}
The Volcano/Cas\-cades framework is well suited for optimizing algebraic expressions, which combine a set of input values using operators to produce an output value. Transformations on an expression generate alternative expressions to compute the same result. The availability of sub-expressions (parts) of an expression is key to Volcano/Cas\-cades, as alternatives for an expression are generated by combining alternatives for sub-expressions (\myOr{} nodes) using operators (\myAnd{} nodes).

However, adapting an algebraic framework such as Volcano/Cas\-cades for optimizing imperative programs is not straight forward. Apart from computing expressions, imperative programs can modify the program stack/heap and contain operations that have side effects (such as writing to a console). Further, real world programs contain complex control and data flow (due to branching, loops, exceptions etc.). 

In this section, we argue that program regions provide a natural abstraction for parts of an imperative program. We then discuss the representation of program alternatives using an \andOrDag{} that we call the \regionDag{}.

\subsection{Region as a State Transition}
\label{sec:programTransition}
An imperative program can be considered as a specification for transition from one state to another. For example, the function \emph{processOrders} from program $P_0$ (Figure~\ref{fig:hib}) specifies the following transition: \emph{by the end} of \emph{processOrders}, variable \emph{result} contains the join of \emph{orders} and \emph{customers} with \emph{myFunc} applied on each tuple. Alternative implementations of the program (such as $P_1$ and $P_2$ from Figure~\ref{fig:alt}) are alternative ways to perform the same transition.

The same argument can be extended to regions. Consider the loop body from program $P_0$ (lines 4 to 6), which is a sequential region. The transition specified by this region is: by the \emph{end of the region}, the contents of the collection \emph{result} at the \emph{beginning of the region} are appended with another element obtained by processing the current tuple. The loop body from program $P_2$ (lines 6 to 8) performs the same computation, however instead of fetching customer information using a separate query as in $P_0$, $P_2$ fetches it from cache.

We now formally define a program/program region as a transition, as follows.
\begin{equation} \label{eq:region_trans}
R: X_0 \rightarrow X_1
\end{equation}
where $R$ is a region, $X_0$ is the state at the beginning of $R$ and $X_1$ is the state at the end of $R$. We call $X_0$ the \emph{input state}, and $X_1$ the \emph{output state}. Since the entire program is also a region, the same definition extends to a program as well.

Our framework is agnostic to the definition of a \emph{state}. For example, in our discussion above, we used the values of program variables (such as \emph{result}) to represent a state. If an application writes to the console, the contents of the console could be included in the definition of state. In general, other definitions may be considered depending on the program transformations used.

For a single statement (basic block), the transition from the input state $X_0$ to the output state $X_1$ involves only the states $X_0$ and $X_1$. For regions that may contain other regions, the transition may involve multiple intermediate states: $(X_0 \rightarrow X_{a1} \rightarrow \ldots \rightarrow X_{an} \rightarrow X_1$) where $X_{a1} \ldots X_{an}$ are results of transitions in sub-regions. The output state of one sub-region feeds as the input state to another sub-region according to the control flow in program.

Our definition of a program region as a transition allows regions to be identified as parts of a program performing local computations that together combine to form the entire program, similar to sub-expressions in an algebraic expression. In this paper, we use the term ``computation in a region $R$'' to refer to the transition from an input state to an output state specified by a region $R$.

\subsection{Region AND-OR DAG}
\label{sec:region_dag}
\begin{figure}
    \centering
    \begin{subfigure}[b]{0.49\linewidth}
        \centering
        \resizebox{0.85\textwidth}{!}{\begin{tikzpicture}[font=\sffamily\large]

\node [andNode, label=0:seq ($P_0.S_{2-7}$)]
(s27)
{};

\node [orNode,below of=s27, xshift=\shiftLeftSmallPlus, yshift=\shiftLeftSmallest]
(b2)
{$P_0.B_2$};
\node [andNode, right of=b2, xshift=\shiftRightSmall, label=0:loop ($P_0.L_{3-7}$)]
(l37)
{};

\node [orNode, below of=l37, xshift=\shiftLeftSmall, yshift=\shiftLeftSmallest]
(b3)
{$P_0.B_3$};
\node [orNode, right of=b3, xshift=\shiftRightSmall]
(s46)
{$P_0.S_{4-6}$};

\node [andNode, below of=s46, yshift=\shiftLeftSmallest, label=west:{seq}]
(s46Op)
{};

\node [noBorderNode, below of=s46Op, yshift=\shiftRightSmallest]
(notShown)
{...};

\path (s27) edge (b2);
\path (s27) edge (l37);
\path (l37) edge (b3);
\path (l37) edge (s46);
\path (s46) edge (s46Op);
\path (s46Op) edge (notShown.north);
\path (s46Op) edge (notShown.north east);
\path (s46Op) edge (notShown.north west);

%
%\path (l1) edge (b1);
%\path (l1) edge (s1);
%
%\path (s1) edge (b2);
%\path (s1) edge (f1);
%
%\path (f1) edge (f11);

\end{tikzpicture}}
        \caption{Region tree}
        \label{fig:region_tree}
    \end{subfigure}
    \begin{subfigure}[b]{0.49\linewidth}
        \centering
        \resizebox{!}{0.75\textwidth}{\begin{tikzpicture}[font=\sffamily\large]

    \node[orNode]
    (s27)
    {$P_0.S_{2-7}$};

    \node [andNode, below of=s27, label=0:seq]
    (s27Op)
    {};

    \node [orNode,below of=s27Op, xshift=\shiftLeftSmallPlus, yshift=\shiftLeftSmallest]
    (b2)
    {$P_0.B_2$};
    \node [orNode,right of=b2, xshift=\shiftRightSmall]
    (l37)
    {$P_0.L_{3-7}$};

    \node [andNode, below of=l37, label=0:loop]
    (l37Op)
    {};

    \node [orNode, below of=l37Op, xshift=\shiftLeftSmall, yshift=\shiftLeftSmallest]
    (b3)
    {$P_0.B_3$};
    \node [orNode, right of=b3, xshift=\shiftRightSmall]
    (s46)
    {$P_0.S_{4-6}$};

    \path (s27) edge (s27Op);
    \path (s27Op) edge (b2);
    \path (s27Op) edge (l37);
    \path (l37) edge (l37Op);
    \path (l37Op) edge (b3);
    \path (l37Op) edge (s46);

\end{tikzpicture}}
        \caption{Initial Region \myDag{}}
        \label{fig:initial_region_dag}
    \end{subfigure}
    
    \vspace{3mm}
    
    \begin{subfigure}[b]{\linewidth}
        \centering
        \resizebox{!}{0.5\textwidth}{\begin{tikzpicture}[font=\sffamily\large]

    \node[orNode]
    (s27)
    {$P_0.S_{2-7}$};

    \node [andNode, below of=s27, label=0:seq]
    (s27Op)
    {};

    \node [orNode,below of=s27Op, xshift=\shiftLeftSmallPlus, yshift=\shiftLeftSmallest]
    (b2)
    {$P_0.B_2$};
    \node [orNode,right of=b2, xshift=\shiftRightSmall]
    (l37)
    {$P_0.L_{3-7}$};

    \node [andNode, below of=l37, yshift=\shiftLeft, label=west:loop]
    (l37Op)
    {};

    \node [orNode, below of=l37Op, xshift=\shiftLeftSmall, yshift=\shiftLeftSmallest]
    (b3)
    {$P_0.B_3$};
    \node [orNode, right of=b3, xshift=\shiftRightSmall]
    (s46)
    {$P_0.S_{4-6}$};

    %alternative P1
    \node [andNode, below of = l37, xshift=\shiftRightBig+\shiftRightSmallest, label=0:seq, label=north:{(1)}]
    (p1s37Op)
    {};
    \node [orNode, below of=p1s37Op, xshift=\shiftLeftSmall]
    (p1b3)
    {$P_1.B_3$};
    \node [orNode, right of=p1b3, xshift=\shiftRightSmall]
    (p1l47)
    {$P_1.L_{4-7}$};

    \node [andNode, below of = p1l47, label=0:loop, yshift=\shiftRightTiny]
    (p1l47Op)
    {};
    \node [orNode, below of=p1l47Op, xshift=\shiftLeftSmall, yshift=\shiftLeftTiny/2]
    (p1b4)
    {$P_1.B_4$};
    \node [orNode, right of=p1b4, xshift=\shiftRightSmall]
    (p1s56)
    {$P_1.S_{5-6}$};

    %alternative P2
    \node [andNode, below of = l37, xshift=\shiftLeftBiggest, label=west:seq, label=north:{(2)}]
    (p2s39Op)
    {};
    \node [orNode, below of=p2s39Op, xshift=\shiftLeftSmall]
    (p2s34)
    {$P_2.S_{3-4}$};
    \node [orNode, right of=p2s34, xshift=\shiftRightSmall]
    (p2l59)
    {$P_2.L_{5-9}$};

    \node [andNode, below of = p2l59, label=0:loop, yshift=\shiftRightTiny, label=west:{(3)}]
    (p2l59Op)
    {};
    \node [orNode, below of=p2l59Op, xshift=\shiftRightSmaller, yshift=\shiftLeftTiny/2]
    (p2s68)
    {$P_2.S_{6-8}$};

    \path (s27) edge (s27Op);
    \path (s27Op) edge (b2);
    \path (s27Op) edge (l37);
    \path (l37) edge (l37Op);
    \path (l37Op) edge (b3);
    \path (l37Op) edge (s46);

    %alternative P1
    \path (p1s37Op) edge (p1b3);
    \path (p1s37Op) edge (p1l47);
    \path [dashed, thick] (l37.south) edge (p1s37Op);
    \path (p1l47) edge (p1l47Op);
    \path (p1l47Op) edge (p1b4);
    \path (p1l47Op) edge (p1s56);

    %alternative P2
    \path [dashed, thick] (l37.south) edge (p2s39Op);
    \path (p2s39Op) edge (p2s34);
    \path (p2s39Op) edge (p2l59);
    \path (p2l59) edge (p2l59Op);
    \path (p2l59Op) edge (p2s68);

    \draw plot [smooth] coordinates  { (p2l59Op.south west) ($ (p2l59Op.south west) + (-2.75em,-1em) $)  ($ (b3.north) + (-1.75em,0.75em) $) (b3.north)};

\end{tikzpicture}}
        \caption{Expanded Region \myDag{}}
        \label{fig:expanded_region_dag}
    \end{subfigure}
    \caption{Representing alternative programs using the \regionDag{}}
    \label{fig:region_dag}
\end{figure}
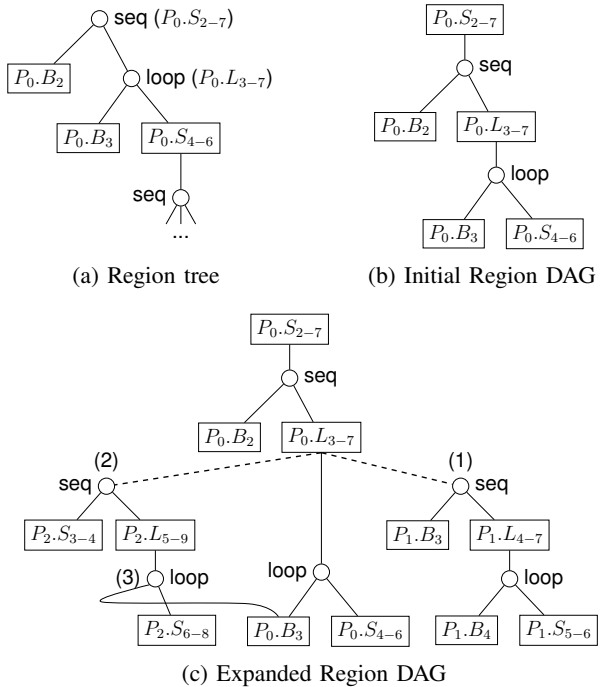
Region \andOrDag{}, or simply \regionDag{}, is an \andOrDag{} that can represent various alternative, but equivalent programs. Given a program with regions, the program and its alternatives can be represented using the Region \myDag{} as follows.\\[6pt]
\textbf{Step 1 \textendash{} Region tree:} Firstly, we identify regions in the program, as described in Section~\ref{subsec:regions}. The hierarchy of regions in a program can be represented as a tree, which we call the \emph{region tree}. The region tree for the regions in Figure~\ref{fig:regions} is shown in Figure~\ref{fig:region_tree}.

The leaves of a region tree are basic block regions. Intermediate nodes are operators that specify how results of sub-regions should be combined to form the parent region. A sequential region is formed using the \emph{seq} operator, a conditional region is formed using the \emph{cond} operator, a loop region using the \emph{loop} operator, and so on. Child nodes are ordered left to right according to the starting line of the corresponding region in the program. In Figure~\ref{fig:region_tree}, we mention the label of the parent region in parentheses along with the operator. The region tree in \adapt{} is analogous to the query expression tree in Volcano/Cas\-cades (Figure~\ref{fig:join_abc_tree}).\\[6pt]
\textbf{Step 2 \textendash{} Initial \regionDag{}:} The next step is to translate the region tree into an \andOrDag{}, which we call the \emph{initial Region \myDag{}}. The initial \regionDag{} for the region tree from Figure~\ref{fig:region_tree} is shown in Figure~\ref{fig:initial_region_dag}. Operator nodes in the region tree are represented as \myAnd{} nodes, and leaf nodes and intermediate results are represented using \myOr{} nodes. The initial \regionDag{} is analogous to the \myDag{} representation of a query in Volcano/Cas\-cades (Figure~\ref{fig:query_and_or_initial}).

An \myOr{} node in the Region \myDag{} represents all alternative ways to perform the computation in a particular region. An \myAnd{} node represents operators to combine sub-regions into the parent region. The initial \regionDag{} contains a single alternative for each region, which is the original program. For example, Figure~\ref{fig:initial_region_dag} represents the following alternative for the region $P_0.S_{2-7}$: perform the computation in the basic block $P_0.B_2$ and then the loop $P_0.L_{3-7}$, sequentially. Similarly, the loop region has a single alternative. Other alternatives may be generated by program transformations.\\[6pt]
\textbf{Step 3 \textendash{} Program transformations}: Program transformations rewrite a program/region to perform the same computation in different ways. In our work, we assume that we are provided with transformations that preserve the equivalence of the original and rewritten programs on any valid input state. \adapt{} then represents these alternative programs efficiently using \regionDag{} for cost based rewriting. Our framework does not infer equivalence of programs or of transformations. It is up to the transformation writer to verify the correctness of transformations. In this paper, we use the transformations from~\cite{EMANI16, RAM12}, with some extensions. We discuss them in Section~\ref{sec:ir}.

In a \regionDag{}, the rewritten program/region is represented as an alternative under the \myOr{} node for that particular region. This may create new nodes in the \regionDag{}. If a node for a region in the rewritten program already exists in the \regionDag{}, it is reused (leveraging techniques in Volcano/Cas\-cades for detecting duplicates and merging nodes). We call the \regionDag{} after adding alternatives from program transformations as the \emph{expanded \regionDag{}}, analogous to the expanded query \myDag{} in Volcano/Cas\-cades (refer Figure~\ref{fig:query_and_or_expanded}).

For example, program transformations such as SQL translation~\cite{EMANI16} and prefetching~\cite{RAM12} identify iterative query invocation inside a loop region in $P_0$, and rewrite the loop as shown in $P_1$ and $P_2$ respectively (refer Figure~\ref{fig:alt}). They are represented in the \regionDag{} as shown in Figure~\ref{fig:expanded_region_dag}. Figure~\ref{fig:expanded_region_dag} shows three alternatives to perform the computation in the loop region $P_0.L_{3-7}$. The newly added alternatives (nodes labeled {\small{\textsf{1}}} and {\small{\textsf{2}}}) are both sequential regions containing a loop region within, and achieve the same result as the original loop region. The loop operator from $P_2$ (node labeled {\small{\textsf{3}}}) shares a basic block ($P_0.B_3$) with the loop region from $P_0$. The loop headers $P_2.B_5$ and $P_0.B_3$ are the same region and the latter already exists in the \regionDag{},  so it is reused.

In summary, there are three alternatives for the root node $P_0.S_{2-7}$, corresponding to the programs $P_0$, $P_1$, and $P_2$. Note that the \andOrDag{} structure allows the node $P_0.B_2$ to be represented only once, although it is part of all three programs corresponding to alternatives for $P_0.L_{3-7}$.

Representing alternative programs in a \regionDag{} is not dependent on an intermediate representation or the program transformations used. Given a program/region and its rewritten version, \adapt{} can represent both the original and transformed programs using the \regionDag{}. This is a key improvement of our representation over Peggy~\cite{peggy}. Peggy aims to represent multiple optimized versions of a program, for the purpose of eliminating the need for ordering compiler optimizations. Representation of programs in Peggy is tied to a specific intermediate representation (IR), which may be provided by the user. Program transformations must be expressed in this IR. \adapt{} on the other hand, does not necessitate the use of an IR, and the transformation process can be unknown to the framework. We present further comparison of our work with Peggy in Section~\ref{sec:relwork}.

Nevertheless, \adapt{} supports representing programs using an IR and expressing transformations on the IR. We discuss one such IR for database applications next, in Section~\ref{sec:ir}. In fact, since the original program is represented intact in the \regionDag{}, it is possible to use multiple IRs simultaneously, each of which may target a specific set of transformations. 

Program regions are essential to representing
alternatives using the Region~\myDag{}. Limitations in the construction
of program regions (discussed in Section~\ref{subsec:regions}) hinder 
the applicability of \adapt{}. For example, in a \emph{try-catch} block, 
control may enter the \emph{catch} block from any statement in the \emph{try} block, 
so it does not conform to the region patterns that we identify. We refer to such fragments with 
complex control flow as \emph{unstructured regions}. Another example of
an unstructured region is an if-else with a complex predicate (combination of two or more
predicates using AND (\&\&) or OR ($||$)), which is 
broken down into simpler predicates by the compiler thereby resulting in complex
control flow. 

Alternatively, these unstructured regions may still be identified 
using a syntactic representation of the program such as an abstract syntax tree (AST).
Unstructured regions may have structured regions within them. For example, 
a \emph{try} block may contain an \emph{if-else} statement. 
In such cases, the unstructured region can be encapsulated into a 
black box, and alternatives can be represented for other parts of 
the program nested within, and outside the unstructured region. 
We omit details.

\section{Transformations using IR}
\label{sec:ir}
\begin{figure}
    \small
    \texttt{\input{snippets/cumulative}}
    \vspace*{0.5em}
    \hrule
    \caption{Program $M_0$: Aggregations inside a loop}
    \label{fig:agg}
\end{figure}

In our earlier work~\cite{EMANI16}, we proposed a DAG based intermediate representation named F-IR (\textit{fold intermediate representation}) for imperative code that may also contain database queries. F-IR has been used to express program transformations for rewriting database applications by pushing relational operations such as selections, projections, joins, and aggregations that are implemented in imperative code to the database using SQL.  In this section, we first present a recap of F-IR from our previous work~\cite{EMANI16}. Later, we describe extensions to F-IR to overcome some of the limitations from~\cite{EMANI16}. We then discuss the integration of F-IR into \adapt{}.

\subsection{F-IR Recap}
\label{sec:fir_recap}
F-IR is based on regions. Variables in a region are represented using expressions only in terms of constants and values available at the beginning of the region; any intermediate assignments are resolved.
F-IR contains operators for representing imperative language operations, as well as relational algebraic operators for representing database queries. Specifically, F-IR uses the \emph{fold} operator (borrowed from functional programming) to algebraically represent loops over col\-lections/query results (which are called \emph{cursor loops}).

For example, consider the program shown in Figure~\ref{fig:agg}, which computes two aggregates, sum and cumulative sum, using a loop over query results. The value of variable \emph{sum} over the loop region is represented with \textit{fold} as follows:\\[3pt]
{\centerline{$\textit{fold}(\textit{\textless{}sum\textgreater{}}+\textit{Q.sale\_amt},0,\textit{Q})$}}\\[3pt]
The first argument to \emph{fold} is the aggregation function. Angular brackets \textless{} and \textgreater{} denote that the value of \emph{sum} in the aggregation function is parametric and is updated in each iteration.
The second argument is the initial value of the aggregate (\emph{sum}) before the loop, in this case $0$ (this feeds as the value of \textless{}\textit{sum}\textgreater{} in the first iteration). The third argument is the query $Q$ over which the loop iterates, in this case: \emph{select month, sale\_amt from sales order by month}. We use the notation $Q.x$ to refer to column $x$ of a tuple in $Q$. Transformations in~\cite{EMANI16} identify the `fold with plus' pattern and infer an SQL query for the variable \emph{sum}, as follows:\\[3pt]
\centerline{{{sum = executeQuery(``\textbf{select} \textbf{sum}(sale\_amt) \textbf{from} sales'');}}}\\

\vspace*{-9pt}
The function \emph{fold} is similar to \textit{reduce} in the map-reduce terminology, and the two functions are referred to synonymously in some contexts. However, there are important differences~\cite{FoldVsReduce} that allow \emph{fold} to represent computations in loops on ordered collections that cannot be represented by \emph{reduce}. For a formal discussion on \emph{fold}, refer~\cite{EMANI16}.

Not all loops can be represented algebraically. We identified in~\cite{EMANI16} the set of preconditions (specified as constraints on inter statement data dependencies) to be satisfied by a cursor loop to represent it using \emph{fold}. However, the preconditions in~\cite{EMANI16} are restrictive as they allow only a single aggregation in a loop to be represented using \emph{fold}. We now discuss this limitation and its impact in the context of cost based transformations. We extend \myFir{} with new operators to overcome the limitation.

\subsection{Extensions to F-IR}
\label{sec:fir_ext}
\begin{figure}
    \centering
    \resizebox {0.85\linewidth} {!} {
        \begin{tikzpicture}[font=\sffamily\large] 
       
    \node [andNode, label=west:{\large{fold}}]
    (fold)
    {};

    \node [andNode, below of=fold, label=west:{\large{tuple}}]
    (idTuple)
    {};

    \node [andNode, below of=fold, xshift=\shiftLeftBiggest, label=west:{\large{tuple}}]
    (functuple)
    {};
    \node [orNode, below of=fold, xshift=\shiftRightBigger]
    (query)
    {{\large{Q}}};

    \node [andNode, below of=functuple, xshift=\shiftLeftSmall, yshift=\shiftLeft - \shiftLeftSmall, label=west:{\large{+}}]
    (agg1)
    {};
    \node [andNode, below of=functuple, xshift=\shiftRightBig, label=west:{\large{map\_put}}]
    (agg2)
    {};

    \node [orNode, below of=idTuple, xshift=\shiftLeftSmaller]
    (cumId)
    {\large{0}};
    \node [orNode, below of=idTuple, xshift=\shiftRightSmall]
    (outputId)
    {\large{\{\}}};

    \node [orNode, below of=agg1, xshift=\shiftLeftSmallPlus]
    (sum)
    {{\large{\textless{}sum\textgreater{}}}};
    \node [orNode, below of=agg1, xshift=\shiftRightSmall]
    (saleAmt)
    {{\large{Q.sale\_amt}}};

    \node [orNode, right of=saleAmt, xshift=\shiftRight]
    (month)
    {{\large{\textless{}cSum\textgreater}}};
    \node [orNode, right of=month, xshift=\shiftRightSmall]
    (colon)
    {{\large{Q.month}}};

    \path (fold.south) edge (functuple);
    \path (fold.south) edge (query.north west);
    \path (fold.south) edge (idTuple);

    \path (functuple) edge (agg1);
    \path (functuple.south east) edge (agg2.north);

    \path (idTuple) edge (cumId);
    \path (idTuple) edge (outputId);

    \path (agg1) edge (sum);
    \path (agg1) edge (saleAmt);

    \path (agg2.south) edge (month);
    \path (agg2.south) edge (colon);

    \draw plot [smooth] coordinates  { (agg2.south) ($ (agg2.south) + (2.5em,-1.5em) $) ($ (agg1.north east) + (3.5em,1em) $) (agg1.north east)};

\end{tikzpicture}
    }
    \caption{F-IR representation for the loop in Figure~\ref{fig:agg}\\
        \emph{Q: select month, sale\_amt from sales order by month}}
    \label{fig:tuple_project}
\end{figure}
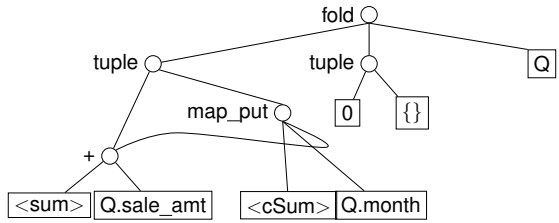

Consider again the program shown in Figure~\ref{fig:agg}. The variable \emph{cSum} cannot be represented in \myFir{} using techniques from~\cite{EMANI16} due to dependent aggregations: i.e., multiple aggregations in a loop, where one aggregate value is dependent on another. In Figure~\ref{fig:agg}, the variable \emph{cSum} is dependent on \emph{sum}.

In our previous work~\cite{EMANI16}, the result of \emph{fold} operator is a single sca\-lar/col\-lection value. When multiple aggregations are present in a loop, we considered separately the part (slice) of the loop computing each aggregation and translated it to SQL separately, as our goal in~\cite{EMANI16} was to translate parts of a program to SQL where possible. For dependent aggregations (such as \emph{cSum} in Figure~\ref{fig:agg}), extracting such a slice is not possible. Thus, the loop cannot be represented as a \emph{fold} expression using techniques from~\cite{EMANI16}. 

An intermediate representation of dependent aggregations in loops is necessary for a cost based decision of transformations. For example, in Figure~\ref{fig:agg}, techniques from~\cite{EMANI16} would extract an SQL query for \emph{sum} (as explained in Section~\ref{sec:fir_recap}) and leave the computation of \emph{cSum} inside the loop intact. Such a rewrite would result in the following program:
\vspace*{6pt}
\hrule
\vspace*{1pt}
\noindent\kw{for}(t : executeQuery(``\kw{select} ... \kw{from} sales \kw{order by} month''))\{\\
\hspace*{0.25cm}sum = sum + t.sale\_amt;\\
\hspace*{0.25cm}cSum.put(month, sum);\\
\}\\
\emph{sum = executeQuery(``\kw{select} \kw{sum}(sale\_amt) \kw{from} sales)};\vspace*{1pt}\hrule
\vspace*{3pt}
However, this transformation degrades program performance, as a new query execution statement (shown in italics) is added to the program resulting in an extra network round trip. Thus, it is necessary in this case that the entire loop be represented in \myFir{} for a cost based decision.

In this paper, we address this limitation by extending the \emph{fold} operator in \myFir{} to return a tuple of expressions. To facilitate this, we introduce two new operators, namely \emph{tuple} and \emph{project}.

\textbf{\emph{tuple}}: The \emph{tuple} operator simply represents a tuple of expressions. It has $n>1$ children, each of which is an expression in \myFir{}. The expressions may have common sub-expressions, which are shared. The output of a \emph{tuple} operator is the n-tuple of outputs of each of its children.

\textbf{\emph{project}}: Intuitively, the \emph{project} operator performs the reverse operation of \emph{tuple}. It takes as input a \emph{tuple} expression and an index $i$, and projects the $i$'th individual expression from \emph{tuple}. In this paper, we represent the index $i$ along with the \emph{project} operator. For example, \emph{project0} projects the first expression from its child \emph{tuple}.

\begin{figure}[!t]
	%\begin{mdframed}
	\textbf{procedure} $\emph{toFIR}(R)$:\hrule \vspace*{3pt}
	{\small
		$R$: A program region.\\
		Let $R.\vartriangle$ be the ee-DAG for $R$, and $R.M$ be its ve-Map.\\
		\textbf{begin}\\
		\hspace*{3pt}\textbf{foreach} sub-region $C$ of $R$\\
		\hspace*{12pt}$\emph{toFIR(C)}$\\
		\hspace*{3pt}\textbf{if} $R$ is not a (cursor) loop region, return\\
		\hspace*{3pt}\textbf{else} $\emph{loopToFold}(R)$\\
		\textbf{end}\\[3pt]}
	\textbf{procedure} $\emph{loopToFold}(R)$:\hrule \vspace*{3pt}
	{\small
		$R$: A cursor loop region\\
		Let $D$ be the data dependence graph for $R$\\
		\textbf{begin}\\
		\textbf{if} there are no external dependency edges in $D$ \{\\
		\hspace*{6pt}Extract the D-IR expression tree $e$ = $\textit{Loop}[Q, e_{\textit{acc}}]$\\
		\hspace*{12pt}for the loop using techniques from~\cite{EMANI16}\\
		\hspace*{6pt}Let $v1_{0}$, $v2_{0}$, $\ldots$ be the initial values of variables \\
		\hspace*{12pt}$v1$,$v2$,$\ldots$ that are updated in the loop.\\
		\hspace*{6pt}\emph{foldExpr} = \emph{fold}$[e _{acc}^{\prime \langle v \rangle, \langle t \rangle},\textit{tuple}(v1_{0}, v2_{0}, \ldots),Q]$\\
		\hspace*{12pt}where, $e _{\textit{acc}} ^{\prime}$ is obtained from $e _{\textit{acc}}$ by 
		replacing each refe-\\
		\hspace*{12pt}rence to attributes of $Q$, with reference to corresponding\\ 
		\hspace*{12pt}attributes of $t$ ($t$ is a new tuple variable).\\
		\hspace*{6pt}\textbf{foreach} variable $v$ that is updated in $R$ \{\\
		\hspace*{12pt}Create the following expression\\
		\hspace*{18pt} \emph{foldExpr}$_v$ = \textit{project}(\textit{foldExpr},$i_v$)\\
		\hspace*{18pt}where $i_v$ is the index of the expression corresponding to \\
		\hspace*{18pt}the variable $v$ in \textit{foldExpr}\\
		\hspace*{12pt}Add \emph{foldExpr} to $R$.$\vartriangle$\\ 
		\hspace*{6pt}\}\\
		\}\\
		\textbf{end}
	}%small end
	%\end{mdframed}
	
	\caption{Algorithm for Conversion to F-IR}
	\label{fig:fir_algo}
\end{figure}

Coupled with \emph{fold}, the operators \emph{tuple} and \emph{project} allow algebraic representation of cursor loops that may have aggregations dependent on one another by removing precondition P2 from~\cite{EMANI16}.  The \myFir{} construction algorithm with modified preconditions is formally shown in Figure~\ref{fig:fir_algo}. The difference of algorithm shown in Figure~\ref{fig:fir_algo} from that in our earlier work~\cite{EMANI16} is that in Figure~\ref{fig:fir_algo}, the precondition restricting the number of aggregated variables in
the loop (labeled P2 in the algorithm from~\cite{EMANI16}) is now removed. For further details, we refer the reader to~\cite{EMANI16}.

Figure~\ref{fig:tuple_project} shows the \myFir{} representation for the loop from Figure~\ref{fig:agg} using \emph{fold}. The aggregation function is a tuple of expressions; one for each aggregated variable (\emph{sum} and \emph{cSum}). Similarly, the initial value passed to \emph{fold} is a tuple that combines the initial values ($0$ and the empty map respectively) for the two aggregates. $Q$ denotes the query from Figure~\ref{fig:agg}. The result of \emph{fold} is a \emph{tuple}. Subsequently, this \myFir{} expression is added to the \regionDag{} for cost based transformations. We discuss this next.

\subsection{Integration into \regionDag{}}
\label{sec:fir_integration}
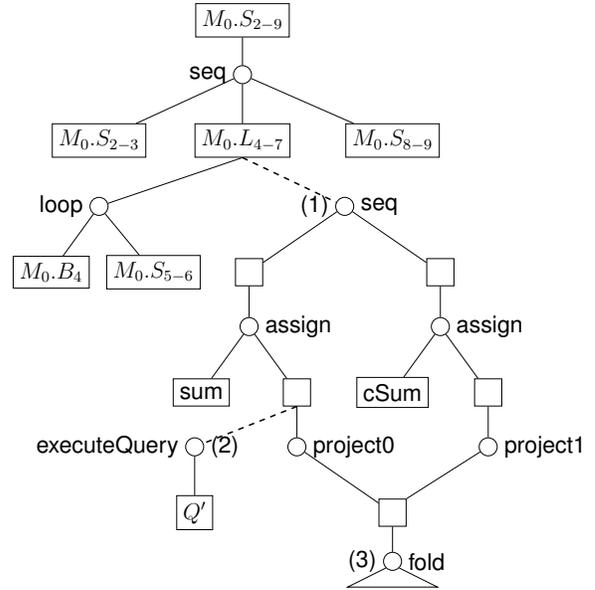
\begin{figure}
    \centering
    \resizebox {0.9\linewidth} {!} {
        \begin{tikzpicture}[font=\sffamily\large]

    \node[orNode]
    (s29)
    {$M_0.S_{2-9}$};

    \node [andNode, below of=s29, label=west:seq]
    (s29Op)
    {};

    \node [orNode,below of=s29Op, xshift=\shiftLeftBig, yshift=\shiftLeftSmallest]
    (s23)
    {$M_0.S_{2-3}$};
    \node [orNode,below of=s29Op, yshift=\shiftLeftSmallest]
    (l47)
    {$M_0.L_{4-7}$};
    \node [orNode,right of=l47, xshift=\shiftRight]
    (s89)
    {$M_0.S_{8-9}$};

    \node [andNode, below of=l47, xshift=\shiftLeftBig, yshift=\shiftLeftSmallest, label=west:loop]
    (l47Op)
    {};
    \node [andNode, right of=l47Op, xshift=\shiftRightBigger, label=0:{seq}, label=west:{(1)}]
    (loopSeqOp)
    {};

    \node [orNode, below of=l47Op, xshift=\shiftLeftSmall, yshift=\shiftLeftSmallest]
    (b4)
    {$M_0.B_4$};
    \node [orNode, right of=b4, xshift=\shiftRightSmall]
    (s56)
    {$M_0.S_{5-6}$};

    %alternative P1
    \node [orNode, below of=loopSeqOp, xshift=\shiftLeft, yshift=\shiftLeftSmallest]
    (sumAssignRoot)
    {};
    \node [orNode, below of=loopSeqOp, xshift=\shiftRight, yshift=\shiftLeftSmallest]
    (cSumAssignRoot)
    {};
    
    \node [andNode,below of=sumAssignRoot, label=0:{assign}]
    (assignSum)
    {};
    \node [andNode,below of=cSumAssignRoot, label=0:{assign}]
    (assignCSum)
    {};
    
    \node [orNode, below of=assignSum, xshift=\shiftLeftSmall, yshift=\shiftLeftSmallest]
    (sum)
    {sum};
    \node [orNode, below of=assignSum, xshift=\shiftRightSmall, yshift=\shiftLeftSmallest]
    (p0)
    {};
    
    \node [orNode, below of=assignCSum, xshift=\shiftLeftSmall, yshift=\shiftLeftSmallest]
    (cSum)
    {cSum};
    \node [orNode, below of=assignCSum, xshift=\shiftRightSmall, yshift=\shiftLeftSmallest]
    (p1)
    {};
    
    \node [andNode, below of=p0, label=0:{project0}]
    (p0Op)
    {};
    \node [andNode, below of=p1, label=0:{project1}]
    (p1Op)
    {};

    \node [orNode, below of=p0Op, xshift=\shiftRight, yshift=\shiftLeftSmallest]
    (fold)
    {};
    \node [andNode, below of=fold, yshift=\shiftRightTiny, label=0:{fold}, label=west:{(3)}]
    (foldOp)
    {};
    
    %query
    \node [andNode,left of=p0Op, xshift=\shiftLeftSmall, label=west:{executeQuery}, label=0:{(2)}]
    (execQuery)
    {};
    \node [orNode,below of=execQuery, yshift=\shiftLeftSmallest]
    (qprime)
    {$Q'$};

    \path (s29) edge (s29Op);
    \path (s29Op) edge (s23);
    \path (s29Op) edge (l47);
    \path (s29Op) edge (s89);
    \path (l47.south) edge (l47Op);
    \path (l47Op) edge (b4);
    \path (l47Op) edge (s56);
%
    %alternative P1
    \path [dashed, thick] (l47.south) edge (loopSeqOp);
    \path (loopSeqOp) edge (sumAssignRoot);
    \path (loopSeqOp) edge (cSumAssignRoot);
    \path (fold) edge (foldOp);
    
    \path (sumAssignRoot) edge (assignSum);
    \path (assignSum) edge (sum);
    \path (assignSum) edge (p0);
    
    \path (cSumAssignRoot) edge (assignCSum);
    \path (assignCSum) edge (cSum);
    \path (assignCSum) edge (p1);
    
    \path (p0) edge (p0Op);
    \path (p1) edge (p1Op);
    
    \path (p0Op) edge (fold);
    \path (p1Op) edge (fold.north east);
    
    %query
    \path [dashed, thick] (p0.south) edge (execQuery);
    \path (execQuery) edge (qprime);
    
    \draw plot coordinates  { (foldOp.south west) ($ (foldOp.south west) + (-2em,-1em) $)  ($ (foldOp.south east) + (2em,-1em) $) (foldOp.south east)};

\end{tikzpicture}
    }
    \caption{\regionDag{} for Figure~\ref{fig:agg} after transforming to \myFir{}
        ($Q'$: \emph{select sum(sale\_amt) from sales}. The \emph{fold} expression 
        (node {\small{\textsf{3}}}) is as shown in Figure~\ref{fig:tuple_project}.)}
    \label{fig:agg_regdag}
\end{figure}

\renewcommand{\arraystretch}{1.2}
\begin{figure*}
    \centering
    \small
    \begin{tabular}{|c|p{8cm}|p{8.5cm}|}
        \hline
        \textbf{Rule} & \hspace*{2cm}\textbf{Definition} & \hspace*{3.5cm}\textbf{Description}\\\hline
        T1 &
        $\textit{fold}(\textit{insert},~\{\},~Q) = Q$  &
        Fold removal (\textit{insert}: set insertion function)\\\hline
        T2 &
        $\textit{fold}(?(\textit{pred},~g),~\textit{id},~Q) \equiv \textit{fold}(g,~\textit{id},\sigma_{\textit{pred}}(Q))$  &
        Predicate push into query (\textit{pred}: predicate; \textit{g}: some function; $?$: conditional execution (if) operator)\\\hline
        T3 &
        $\textit{fold}(g(v,~h(Q.A)),~\textit{id},~Q) \equiv \textit{fold}(g,~\textit{id},\pi_{h(A)}(Q))$  &
        Push scalar functions into query (\textit{g},\textit{h}: functions; \textit{A}: column in $Q$)\\\hline
        T4 &
        $\textit{fold}(\textit{fold}(\textit{insert},~\textit{id},~\sigma_{\textit{pred}}(Q_2)),~\{\},~Q_1) \equiv Q_1 \Join_{pred} Q_2$  &
        Join identification (\textit{pred}: a predicate; \textit{insert}: set insertion function)\\\hline
        T5 &
        $\textit{fold}(\textit{op},~\textit{id},~\pi_{A}(Q)) \equiv \gamma_{\textit{op\_agg}(A)}(Q)$  &
        Aggregation (\textit{op}: a binary operation like +, scalar \textit{max}; \textit{op\_agg}: corresponding relational aggregation operation like \textit{sum}, \textit{max})\\\hline
        N1 &
        $\textit{fold}(\textit{f}(v,\textit{executeQuery}(\sigma_{R.A=Q.B}(R))),\textit{id},Q) \equiv \textit{seq}(\textit{prefetch}(R,A),\textit{fold}(\textit{f}(v,\textit{lookup}(Q.B)),\textit{id},Q))$  &
        Prefetching (\textit{prefetch}: fetch query result and cache by column locally. \textit{cacheByColumn, lookup}: Refer footnote \ref{footnote:cache}). \\\hline
        N2 &
        $ \textit{fold}(g,~\textit{id},\sigma_{\textit{pred}}(Q)) \equiv \textit{fold}(?(\textit{pred},g),\textit{id},Q)$ &
        Reverse of T2\\\hline
    \end{tabular}
    \caption{F-IR Transformation Rules (T1 to T5 are from~\cite{EMANI16})}
    \label{fig:ir_trans}
\end{figure*}
\renewcommand{\arraystretch}{1}

As we mentioned earlier in Section~\ref{sec:ir}, \myFir{} is based on regions, and \myFir{} expressions represent values of program variables at the end of a region in terms of values available at the beginning of a region. Thus, an \myFir{} expression also specifies a transition from an input state to an output state in a region, where the input and output states consist of values of all program variables that are live at the beginning and at the end of the region, respectively.

We model the construction of an \myFir{} expression for a region as a program transformation that takes a region as input and gives the equivalent \myFir{} expression as output. If the preconditions for \myFir{} representation (Section~\ref{sec:fir_ext}) are satisfied, the \myFir{} expression is constructed and added as an alternative to the corresponding region. If the preconditions fail, no \myFir{} expressions are added, but other program transformations can still be applied on the \regionDag{}. 

Figure~\ref{fig:agg_regdag} shows the \regionDag{} for program $M_0$ from Figure~\ref{fig:agg}. The program consists of a sequential region ($M_0.S_{2-9}$) containing a loop region within ($M_0.L_{3-6}$). The \myFir{} expression from Figure~\ref{fig:tuple_project} is used to add an alternative (node {\small{\textsf{1}}}) to the loop region. Using the \emph{fold} expression for the loop, we first extract the individual variable values using \emph{project}, assign them to the appropriate variables, combine the assignments using a \emph{seq} operator, and add the alternative to the \myOr{} node corresponding to the loop. 

\subsection{Transformations}
\label{sec:fir_trans}
\review{Transformations on \myFir{} expressions add further alternatives to the \regionDag{}. In our earlier work~\cite{EMANI16}, we proposed \myFir{} transformations with the aim of translating imperative code into SQL. These transformations are summarized in Figure~\ref{fig:ir_trans} (T1 to T5)\footnote{%
    $\gamma$ is the relational aggregation operator. Here, we present abridged versions of the rules, for the sake of brevity. For complete details of these transformations including ordering, duplicates, and variations of each rule, refer~\cite{EMANI16}.}%
. (There are other transformation rules in~\cite{EMANI16}, all of which are included in our implementation.) Prefetching is widely used in enterprise settings to mitigate the cost of multiple invocations of the same query. To enable prefetching, in this paper, we propose new transformations N1 and N2 (Figure~\ref{fig:ir_trans}). Rule N1 transforms iterative lookup queries inside a loop into a prefetch\footnote{%
In our current implementation, N1 prefetches an entire relation and all subsequent lookups are performed locally. This can be extended to prefetch queries that result only in a part of the relation.
}
followed by local cache lookups. Rule N2 transforms a selection query into a query without selection followed by a local filter. 
Note that rule N1 uses a combination of \myFir{} operators as well as operators for combining regions (such as \emph{seq}, \emph{loop} and \emph{cond}).}

\review{We use Rule T5 to extract an SQL query for \emph{sum}. This is added as an alternative (node {\small{\textsf{2}}}) to the \myOr{} corresponding to the expression for \emph{sum}. Similarly, alternative expressions for \emph{cSum} are added after applying other transformations. Using the cost model described in Section~\ref{sec:costing}, \adapt{} can identify that the alternative with node {\small{\textsf{2}}} incurs an extra query execution cost, in addition to the loop computation represented by \emph{fold}. After the least cost program is found, the \myFir{} representation is translated into imperative code. We refer the reader to~\cite{EMANI16} for details on generating imperative code from \myFir{}.}\\[3pt]
\noindent
\review{\textbf{Limitations of \myFir{}}: As discussed earlier (Section~\ref{sec:fir_recap}), not all loops can be represented in \myFir{}. The focus of \myFir{} is to represent set-oriented operations on collections/query results using cursor loops in imperative programs. Further, \myFir{} currently represents only selection (read) queries, so updates are not part of \myFir{}. Expanding \myFir{} to support updates is part of future work. Note that selection queries interleaved with update queries can still be represented using \myFir{}, leaving the updates intact. We refer the reader to~\cite{EMANI16} for more details.}

\section{Cost Model}
\label{sec:costing}
In this section, we discuss how to estimate the cost of a program represented using the \regionDag{}, and how to find the best alternative from many possible alternatives. We will restrict our attention to cost estimation for individual nodes in the \regionDag{}; the idea for cost based search in the \regionDag{} is similar to that in the Volcano/Cas\-cades \andOrDag{} (refer Section~\ref{sec:volcano_and_or}).
\begin{figure}
\begin{center}
    \small
    \begin{tabular}{|c|p{7cm}|}
        \hline
        \textbf{Term} & \hspace*{2cm}\textbf{Definition} \\\hline
        $C_{\mli{NRT}}$ & Network round trip time between the client (where the program is running) and the database.\\\hline
        \review{$C_{Q}^F$} & \review{Time taken by the database since receiving the query to send out the first row in the result.}\\\hline
        \review{$C_{Q}^L$} & \review{Time taken by the database since receiving the query to send out the last row in the result.}\\\hline
        $N_Q$ &  Cardinality of the result set for Q, i.e., the number of rows in the result after executing Q.\\\hline
        $S_{row(Q)}$ &  Size in bytes of a single row in the result set for Q.\\\hline
        \emph{BW} &  Network bandwidth (bytes/sec)\\\hline
        \review{$\emph{AF}_Q$} & \review{Amortization factor \textendash{} estimated number of invocations of Q.}\\\hline
        $C_Y$ & Cost of a program operator node in the \regionDag{}\\\hline
        $C_Z$ & Cost of executing one imperative program statement (other than query execution statement)\\\hline
    \end{tabular}
\end{center}
\caption{Cost parameters}
\label{fig:cost_params}
\end{figure}

In our work we focus on optimizing programs for data access.
Figure~\ref{fig:cost_params} describes the parameters we 
consider for cost estimation. We use a parameter 
\textit{amor\-ti\-zation factor} (\textit{AF}$_Q$) that estimates the 
number of invocations of a query Q, to allocate the prefetching
cost across each invocation. 

Determining whether or not a relation should be prefetched 
is non trivial, as this may affect the cost of other nodes
included in a plan. This problem is similar to the multi-query 
optimization problem, which aims to calculate the best cost
and plan for a query considering materialization~\cite{ROY00} 
(in our case, caching). Currently in our framework, we decide to prefetch a query
if (a) it is explicitly marked for prefetching as the result of a
transformation (such as N1 from Figure~\ref{fig:ir_trans}), or (b)
an entire relation is fetched without any fil\-ters/group\-ing. AF may 
be tuned individually for various queries 
depending on the particular application's workload.

We note however, that using prefetching, the first access to the query
may have significantly higher latency compared to the original program,
as typically a large number of rows are prefetched using a single query.
This can be mitigated by prefetching asynchronously, and dynamically
deciding to prefetch only after a certain number of accesses to minimize 
the overhead of prefetching. This is similar to the classical ski-rental 
problem~\cite{skiRental88} and has been applied earlier in the context of
join optimizations in parallel data management systems~\cite{Chandra17}. 
Extending \adapt{}
to adapt heuristics from~\cite{ROY00} to efficiently handle
alternatives generated due to caching is part of future work, and
dynamic approaches for prefetching are part of future work.

Currently, we calculate cost only in terms of the 
time taken to execute the program. Our cost model 
can be extended to include other parameters such as 
CPU cost, memory usage etc., if needed. Using the parameters 
from the table above, the cost of various nodes in the 
\andOrDag{} is estimated as follows.\\[3pt]
\noindent
\review{\textbf{\emph{Query execution}}: The cost of execution of a query Q is defined as follows:\\
\centerline{$C_Q$ = $C_{\mli{NRT}}$ + $C_{Q}^F$ + \emph{max}($N_Q$*$S_{row(Q)}/\textit{BW}$, $C_{Q}^L-C_{Q}^F$)}\\[3pt]
\noindent
\textbf{\emph{Prefetch}}: The cost of prefetching a relation using a query Q is defined as follows:\\
\centerline{$C_{\textit{prefetch}(Q)}$ = $C_Q / \textit{AF}_Q$}}\\[3pt]
\noindent
\textbf{\emph{Basic block node}}: A basic block node in the \regionDag{} contains imperative code. The cost of the basic block is the sum of the cost of each statement ($C_Z$) in the basic block. $C_Z$ can be tuned according to the particular application. \eat{Currently, we do not distinguish between the different types of imperative statements, except for query execution statements.}\\[3pt]
\noindent
\textbf{\emph{Region operator node}}: Region operator nodes are rooted at the operators \emph{seq}, \emph{cond}, or \emph{loop}. Their cost is calculated as follows:\\
$C_\mli{seq}$ = sum of cost of each child.\\[3pt]
$C_\mli{cond}$ = \emph{p}~*~$C_\mli{true}$ + \emph{(1-p)}~*~$C_\mli{false}$ + $C_{p}$\\ where \emph{p} is the probability that the condition evaluates to true, $C_p$ is the cost of evaluating the condition, and $C_\mli{true}$ and $C_\mli{false}$ are the costs of the sub regions corresponding to \emph{p} evaluating to true and false respectively. If the condition is in terms of a query result attribute, our framework estimates the value of \emph{p} using database statistics. Otherwise, a value of 0.5 is used.\\[3pt]
$C_\mli{loop}$: If the loop is over the results of a query Q, then it may be represented using a \emph{fold} expression, whose cost is calculated as follows:\\
\centerline{$C_\mli{fold}$ = $N_Q$ * $C_f$ + $C_{\mli{Db(Q)}}$}\\
where $C_f$ is the cost of the fold aggregation function.

If the number of iterations is known (loop is over the results of a query, or over a collection) but the loop cannot be represented using \emph{fold}, then the cost is calculated as \emph{K} * $C_\mli{body}$, where $C_\mli{body}$ is the cost of the loop body, and \emph{K} is the number of loop iterations. If the number of iterations cannot be known (such as in a generic while loop), we use an approximation for the number of loop iterations, which can be tuned according to the application.\\[3pt]
\noindent
\textbf{\emph{Other \myFir{} operators}}: We assign a static cost $C_Y$ for evaluating any other \myFir{} operator. $C_Y$ can be tuned according to the particular application.
\eat{Sometimes, the cost of a particular node may depend on other nodes in the \andOrDag{} apart from its children due to side effects such as caching, as multiple nodes can access results from the cache without incurring further execution costs for the same query. In our current implementation, we cache all query results (which are typically small in ORM applications) in memory and reuse them, thus incurring the cost of query execution only once. In general, this is similar to the problem of multi-query optimization with materialization. Greedy heuristics for multi-query optimization proposed in~\cite{ROY00} can be adapted to efficiently handle alternatives generated due to caching of multiple queries with large results. Implementing this in \adapt{} is an area of future work.}

\section{Related Work}
\label{sec:relwork}
In this section, we survey related work on various fronts.\\[3pt]
\noindent
\review{\textbf{Program transformations for database applications}: In our earlier work, we have developed the DBridge system~\cite{DBR, DBR17} for optimizing database applications using static program analysis techniques. Various program transformations such as batching, asynchronous query submission and prefetching~\cite{RAM12,TKDE14} have been incorporated in DBridge. DBridge also contains transformations for rewriting Hibernate applications using SQL for improved performance~\cite{EMANI16}; the QBS system~\cite{Cheung13} also addresses the same problem. However, existing approaches assume that such transformations are always beneficial. In contrast, our framework allows a cost-based choice of whether or not to perform a transformation, and to choose the least cost alternative from more than one possible rewrites.}

\review{Note that unlike earlier techniques in DBridge~\cite{RAM12, TKDE14, EMANI16}, the focus of this paper is not on the program trans\-for\-mations themselves; rather we focus on representing various alternatives produced by one or more transformations of imperative code and choosing the least cost alternative. Our implementation of \adapt{} uses DBridge as a sub-system for generating alternative programs by applying these transformations. In general, \adapt{} can be used independent of DBridge with any set of program transformations.}

\review{There has been work on automatically rewriting programs
with embedded queries for evolving schemas, using program 
transformations that are derived from schema modifications~\cite{shneiderman1982}.
The transformations we considered in our work instead focus on
rewriting queries for a fixed schema, by pushing computation 
from imperative code into SQL. However, \adapt{} can be used for
cost based rewriting of applications using transformations 
from~\cite{shneiderman1982}.}\\[3pt]
\noindent
\textbf{Enumeration and application of transformations:} The Peggy compiler optimization framework~\cite{peggy} facilitates the application of transformations (compiler optimizations) in any order. It uses a data structure called PEG that operates similar to the Volcano/Cas\-cades \andOrDag{}. However, there are significant differences from our framework.

Peggy is aimed at compiler optimizations and works on expressions. Our framework is aimed at transformations on larger program units such as regions or even an entire program in addition to transformations on expressions, and can support multiple IRs unlike Peggy (as discussed in Section~\ref{sec:and_or_rep}). \adapt{} also improves upon Peggy in terms of program cost estimation. The cost model in Peggy is primitive, especially as the cost of a loop is calculated as a function of its nesting level and a predetermined constant number of iterations. Such a cost model is inadequate for database applications as query execution statements and loops over query results take the bulk of program execution time. A more sophisticated cost model that can use the database and network statistics, such as the one described in this paper, is desired.\\[3pt]
\noindent
\textbf{Pushing computation to the database:} The Pyxis~\cite{Cheung2012} system automatically partitions database applications so that a part of the application code runs on a co-located JVM at the database server, and another part at the client. In contrast to Pyxis, \adapt{} generates complete and equivalent programs using program transformations on the original program, and does not require any special software at the database server.\\[3pt]
\noindent
\review{\textbf{LINQ to SQL:} A number of language integrated querying 
frameworks similar to LINQ~\cite{LINQ} allow developers to express 
relational database queries using the same language as the application,
and later translate these queries into SQL~\cite{LINQ, grust2010}. 
Our techniques focus on automatically identifying parts
of imperative code that can be pushed into SQL, 
whereas \cite{grust2010} require developers to completely 
specify these queries, albeit in a syntax that uses 
source language constructs.}
%

%\balance
\section{Experimental Evaluation}
\label{sec:experiments}
\begin{figure*}
	\centering
	\begin{subfigure}[b]{0.32\linewidth}
		\centering
		\resizebox{!}{0.88\textwidth}{\begin{tikzpicture}
\begin{axis}[
xlabel={\large{No. of \emph{Orders} rows (log scale)}},
xmode=log,
ymode=log,
xmin=100,
xmax=1000000,
ymin=1,
ymax=10000,
xtick={100,1000,10000,100000,1000000},
xticklabels={$100$,$1$km,$10$k,$100$k,$1$m},
ytick={1,10,100,1000,10000},
yticklabels={$1$,$10$,$100$,$1$k,$10$k},
extra y ticks={3467, 6047},
extra y tick labels={$3467$,$6047$},
extra y tick style={yticklabel pos=right,ytick pos=right},
ymajorgrids=true,
xmajorgrids=true,
grid style=dashed,
legend style={draw=none, at={(0.95,0.05)},anchor=south east},
legend cell align={left},
legend entries = {Hibernate(P0), SQL Query(P1), Prefetching(P2), \adapt{}},
]

%P0
\addplot[color=black,very thick,mark=x,mark size=3pt]
coordinates {(100,29.627)(1000,272.393)(10000,2523.66)(100000,25023)(1000000,250023)};

%P1
\addplot[color=darkgreen,very thick,mark=square,mark size=3pt]
coordinates {(100,3.744)(1000,12.95)(10000,66.147)(100000,620.163)(1000000,6047.49)};

%P2
\addplot[color=red,very thick,mark=diamond, mark size=3pt]
coordinates {(100,229.207)(1000,230.236)(10000,252.802)(100000,570.624)(1000000,3467.56)};

%Cobra
\addplot[color=blue,line width=2pt,mark=+, mark size=3pt, mark options={solid}, loosely dashdotted,]
coordinates {(100,3.744)(1000,12.95)(10000,66.147)(100000,570.624)(1000000,3467.56)};

\end{axis}

\node[rotate=90,anchor=center] at ([xshift=-3pt]current bounding box.west) {\hspace*{1cm}\large{Prog. execution time(s, log scale)}};

\end{tikzpicture}}
		\caption{Slow remote nw - varying Orders}
		\label{fig:low}
	\end{subfigure}
	\hspace*{7pt}
	\begin{subfigure}[b]{0.32\linewidth}
		\centering
		\resizebox{!}{0.85\textwidth}{\begin{tikzpicture}
\begin{axis}[
xlabel={\large{No. of \emph{Orders} rows (log scale)}},
xmode=log,
ymode=log,
xmin=100,
xmax=1000000,
ymin=0.1,
ymax=30,
xtick={100,1000,10000,100000,1000000},
xticklabels={$100$,$1$k,$10$k,$100$k,$1$m},
ytick={0.1,1,10},
yticklabels={$0.1$,$1$,$10$},
extra y ticks={12, 16},
extra y tick labels={$12$,$16$},
extra y tick style={yticklabel pos=right,ytick pos=right},
ymajorgrids=true,
xmajorgrids=true,
grid style=dashed,
legend style={draw=none, at={(1,0)},anchor=south east},
legend cell align={left},
legend entries = {Hibernate(P0), SQL Query(P1), Prefetching(P2), \adapt{}},
legend pos = north west
]

%P0
\addplot[color=black,very thick,mark=x,mark size=3pt]
coordinates {(100,0.2204)(1000,0.5562)(10000,1.7988)(100000,7.0294)(1000000,18.1852)};

%P1
\addplot[color=darkgreen,very thick,mark=square,mark size=3pt]
coordinates {(100,0.1892)(1000,0.2984)(10000,0.6196)(100000,2.0918)(1000000,16.0746)};

%P2
\addplot[color=red,very thick,mark=diamond, mark size=3pt]
coordinates {(100,0.9088)(1000,0.9554)(10000,1.0592)(100000,1.841)(1000000,11.7632)};

%Cobra
\addplot[color=blue,line width=2pt,mark=+, mark size=3pt, mark options={solid}, loosely dashdotted,]
coordinates {(100,0.1892)(1000,0.2984)(10000,0.6196)(100000,1.841)(1000000,11.7632)};

\end{axis}

\end{tikzpicture}}
		\caption{Fast local nw - varying Orders}
		\label{fig:high}
	\end{subfigure}
	\hspace*{-4pt}
	\begin{subfigure}[b]{0.32\linewidth}
		\centering
		\resizebox{!}{0.85\textwidth}{\begin{tikzpicture}
\begin{axis}[
xlabel={\large{No. of \emph{Customers} rows (log scale)}},
xmode=log,
ymode=log,
xmin=10,
xmax=100000,
ymin=10,
ymax=400,
xtick={10,100,1000,10000,100000},
xticklabels={$10$,$100$,$1$k,$10$k,$100$k},
ytick={10,100},
yticklabels={$10$,$100$},
extra y ticks={30,300},
extra y tick labels={$30$,$300$},
extra y tick style={yticklabel pos=left,ytick pos=left},
ymajorgrids=true,
xmajorgrids=true,
grid style=dashed,
legend style={draw=none, fill=none, at={(0.975,0.025)},anchor=south east},
legend cell align={left},
legend entries = {Hibernate(P0), SQL Query(P1), Prefetching(P2), \adapt{}},
]

%P0
\addplot[color=black,very thick,mark=x,mark size=3pt]
coordinates {(10,33.2075)(100,57.157)(1000,295.42)(10000,1000)(100000,10000)};

%P1
\addplot[color=darkgreen,very thick,mark=square,mark size=3pt]
coordinates {(10,56.251)(100,56.389)(1000,56.787)(10000,57.236)(100000,57.728)};

%P2
\addplot[color=red,very thick,mark=diamond, mark size=3pt]
coordinates {(10,30.748)(100,31.408)(1000,33.903)(10000,58.199)(100000,301.683)};

%Cobra
\addplot[color=blue,line width=2pt,mark=+, mark size=3pt, mark options={solid}, loosely dashdotted,]
coordinates {(10,30.748)(100,31.408)(1000,33.903)(10000,57.236)(100000,57.728)};

\end{axis}

\node[rotate=90,anchor=center] at ([xshift=3pt]current bounding box.east) {\hspace*{0.8cm}\large{Prog. execution time(s, log scale)}};

\end{tikzpicture}}
		\caption{Slow remote nw - varying Customers}
		\label{fig:low_bw_varying_cust}
	\end{subfigure}

	\caption{Performance of alternative implementations of Figure~\ref{fig:hib} in slow remote and fast local networks}
	\label{fig:variations}
\end{figure*}
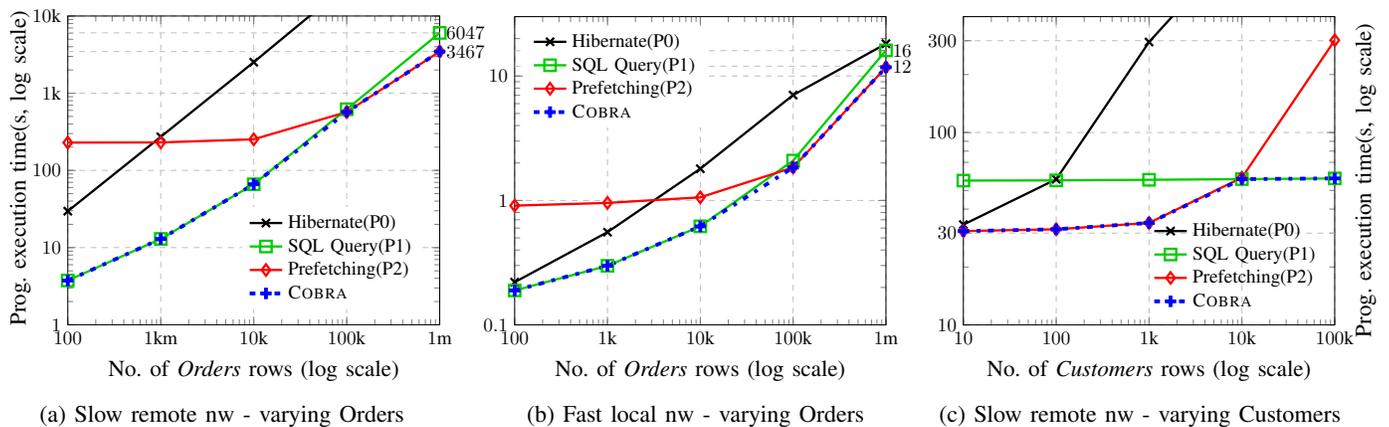

In this section, we present an evaluation of the \adapt{} framework for cost based rewriting of database applications. We implemented \adapt{} by extending the PyroJ optimizer~\cite{ROY00}, which is based on Volcano/Cas\-cades. \adapt{} leverages the region based analysis framework and program transformations from the DBridge~\cite{DBR17} system for optimizing database applications. DBridge internally uses the Soot framework~\cite{soot} for static analysis.

For our experiments, we used two machines: a server that runs the database (16GB RAM with Intel Core i7-3770, 3.40GHz CPU running MySQL 5.7 on Windows 10), and a client that runs the application programs (8GB RAM with Intel Core i5-6300 2.4GHz CPU running Windows 10, around 4GB RAM was available to the application program). The numbers reported in the experiments are averaged over five runs of the program. 

Our experiments aim to evaluate the following: (a) applicability of \adapt{} and our cost model and (b) performance benefits due to cost based rewriting. Our experiments use real world and synthetic code samples that use the Hibernate ORM.

In Experiments 1, 2, and 3, we evaluate the performance of program $P_0$ and its alternatives $P_1$ and $P_2$ (which were shown in Figure~\ref{fig:alt}), along with the choice suggested by \adapt{}. We implemented $P_0$ using the Hibernate ORM, and used transformation rule N1 and a variation of transformation rule T5 (refer Section~\ref{sec:fir_integration}) to generate $P_2$ and $P_1$ respectively, from $P_0$. The size of each row in \emph{Order} and \emph{Customer} has been chosen according to the TPC-DS~\cite{TPCDS} benchmark specification.

We ran the programs under varying network conditions and cardinalities of the tables \emph{Order} and \emph{Customer}. We connected the client and server directly with an ethernet cable, and simulated variations in the network using a network simulator~\cite{NEWT}. We used the following conditions: \emph{slow remote network} (bandwidth: 500kbps, latency: 250ms~(taken from~\cite{awsmap})) and \emph{fast local network} (bandwidth: 6gbps, round trip time: 0.5ms).

For estimating the cost of generated alternatives using our cost model, we focused on data transfer costs and number of loop iterations (size of query result set). \review{The cost of executing any other instruction apart from a query execution statement in the imperative program ($C_z$ from Section~\ref{sec:costing}) was set to 30ns, after profiling the applications to estimate the same. We set the amortization factor to 1 (for experiments 1, 2 and 3). We consulted the database query optimizer to get an estimate of query execution times, based on past executions of the queries.}
The cost metrics we used were provided to our system as a cost catalog file.\\[3pt]
\noindent
\textbf{Experiment 1:} We first ran the programs using a slow remote network. We fixed the number of rows in \emph{Customer} to 73,000 and varied the number of \emph{Order} rows from 100 to 1 million. Figure~\ref{fig:low} shows the actual running times of these programs, and the choice suggested by \adapt{}. At lower number of \emph{Order} rows, \adapt{} chose the program using SQL query API ($P_1$), as the other two alternatives incur high latency. Program $P_0$ suffers from large number of network round trips due to iterative queries, and $P_2$ prefetches a relatively large amount of \emph{Customer} data. However, as the number of \emph{Order} rows approaches the number of \emph{Customer} rows, program $P_1$ causes increasing duplication of \emph{Customer} data in the join result. At this point, \adapt{} switched to program $P_2$. The performance of prefetching ($P_2$) does not vary much for lower cardinalities as the bulk of the time is spent on fetching the larger relation (\emph{Customer}) data. In each case, \adapt{} correctly identified the least cost alternative.\\[3pt]
\noindent
\textbf{Experiment 2:} We use the same cardinalities as in Experiment 1, but use a fast local network. Again, \adapt{} estimated $P_1$ to be the least cost alternative until the number of \emph{Order} rows approaches the number of \emph{Customer} rows, and switched to $P_2$ after that. This is reflected in the running times of these programs, as shown in Figure~\ref{fig:high}. Although $P_2$ performs better than $P_1$ at high cardinality of \emph{Order} in both Figure~\ref{fig:high} and Figure~\ref{fig:low}, the performance difference is much more significant in a slow remote network (3467s vs 6047s) than in a fast local network (12s vs 16s). Note that the performance of SQL query ($P_1$) and Hibernate ($P_0$) is comparable at high cardinalities in fast local network. This can be understood as follows. The overhead of a network round trip is very small in a fast local network. Hibernate program internally caches each \emph{Customer} row once fetched, so the latency is minimized after all \emph{Customer} rows have been fetched using individual queries.\\[3pt]
\noindent
\textbf{Experiment 3:} In this experiment,  we use a slow remote network, fix the number of \emph{Order} at 10,000 and vary the number of \emph{Customer} rows. As the results from Figure~\ref{fig:low_bw_varying_cust} indicate, the time taken by $P_1$ is nearly constant (as the size of the join result does not vary with increasing number of \emph{Customer} rows). However, the time taken by $P_2$ increases with the number of \emph{Customer} rows as $P_2$ prefetches the entire \emph{Customer} table. This demonstrates that unlike Figures~\ref{fig:high} and ~\ref{fig:low}, it is not necessary that $P_1$ performs better at lower cardinalities, and $P_2$ performs better at higher cardinalities. \adapt{} correctly chose the least cost program in each case based on its cost model.\\[3pt]
\begin{figure}
    \centering
    \small
    \begin{tabular}{|c|p{7cm}|c|}
        \hline
        \footnotesize{\textbf{Id}} &\hspace*{1cm}\footnotesize{\textbf{Description of cost based choice}} &\footnotesize{\textbf{\#}}\\\hline
        A &
        \textit{Nested loops with intermittent updates}: Inner loop can be translated to SQL for better performance \underline{vs} overall performance may degrade due to iterative queries
          &
        3\\\hline
        B &
        \textit{Multiple aggregations inside loop}: Faster aggregation/fetch only result by translation to SQL \underline{vs} multiple queries (NRT) instead of one &
        2\\\hline
        C &
        \textit{Nested loops join}: Better join algo. at the database and fetch (large) result of SQL join \underline{vs} Cache tables at application and join locally &
        9\\\hline
        D &
        \textit{Function that is called inside a loop can be rewritten using SQL}: overall performance may degrade due to iterative queries if caller loop cannot be translated&
        7\\\hline
        E &
        \textit{Collection filtered differently across different calls of a recursive function}: Multiple point look up queries \underline{vs} prefetch whole table once and filter from cache
        &
        9\\\hline
        F &
        \textit{Different parts of a collection are used across different callee functions}: Multiple select/project queries to fetch only required data \underline{vs} prefetch all data with one query  &
        2 \\\hline
    \end{tabular}
    \caption{Cases for cost based based optimization in real world application (pattern id, description, number of cases)}
    \label{fig:cases}
\end{figure}
\noindent
\textbf{Experiment 4:} In this experiment, we used a real world open source application, Wilos~\cite{Wilos}, which uses the Hibernate ORM framework. By manual examination of the Wilos source code, we identified 32 code samples where cost based transformations are applicable. These samples can be broadly classified into six categories. Figure~\ref{fig:cases} lists for each category, the cost based choice of transformations and the number of cases identified. Details of each code fragment are listed in Appendix~\ref{sec:cases_details}. 

\begin{figure}
	\centering
	\resizebox{0.45\textwidth}{!}{\begin{tikzpicture}[
every axis/.style={ % add these settings to all the axis environments in the tikzpicture
	ybar stacked,
	xlabel={\small{Program ID}},
	ymin=0,
	ymax=1.5,
	symbolic x coords={P\_A,P\_B,P\_C,P\_D,P\_E,P\_F},
	bar width=6pt,
	width=0.55\textwidth,
	height=0.3\textwidth,
	xtick style={draw=none},
	ytick={0.2,0.4,0.6,0.8,1,1.2,1.4},
	yticklabels={\small{$0.2$},\small{$0.4$},\small{$0.6$},\small{$0.8$},\small{$1$},\small{$1.2$},\small{$1.4$}},
	legend style={draw=none,font=\footnotesize},
	legend cell align={left},
	point meta=explicit symbolic,
	nodes near coords,
},
]

% bar shift -10pt here
\begin{axis}[bar shift=-7pt,hide axis,legend style={at={(0.35,0.99)},anchor=north west,draw=none, fill=none}]
\addplot[black,fill=none,postaction={pattern=crosshatch, pattern color=black}] coordinates
{(P\_A,1)[\textcolor{black}{\hspace*{-20pt}\scriptsize{$3.9$s}}] 
	(P\_B,1)[\textcolor{black}{\hspace*{-24pt}\scriptsize{$0.9$s}}]
	(P\_C,1)[\textcolor{black}{\hspace*{-14pt}\scriptsize{$6087$s}}] 
	(P\_D,1)[\textcolor{black}{\hspace*{-14pt}\scriptsize{$0.6$s}}]
	(P\_E,1)[\textcolor{black}{\hspace*{-14pt}\scriptsize{$1.9$s}}] 
	(P\_F,1)[\textcolor{black}{\hspace*{-14pt}\scriptsize{$2.1$s}}]
};
\addlegendentry{Original};
\addplot coordinates
{(P\_A,0) (P\_B,0) (P\_C,0) (P\_D,0) (P\_E,0) (P\_F,0)};
\end{axis}

% zero bar shift here    
\begin{axis}[hide axis, legend style={at={(0.35,0.89)},anchor=north west, draw=none, fill=none}]
\addplot+[red,fill=none,postaction={pattern=north east lines, pattern color=red}] coordinates
{(P\_A,1.04) (P\_B,1.327) (P\_C,0.006)[\textcolor{black}{\hspace*{12pt}\scriptsize{$42.5$s}}] (P\_D,0.754) (P\_E,1) (P\_F,0.556)};
\addlegendentry{Heuristic};
\addplot+[fill=none] coordinates
{(P\_A,0) (P\_B,0) (P\_C,0) (P\_D,0) (P\_E,0) (P\_F,0)};
\end{axis}

% and bar shift +10pt here
\begin{axis}[bar shift=7pt,legend entries={\adapt{}${(\textit{AF=}50)}$, \adapt{}${(\textit{AF=}1)}$}, legend style={at={(0.55,0.99)},anchor=north west,draw=none, fill=none}]
\addplot+[black,fill=green!50!gray] coordinates
{(P\_A,0.062066) (P\_B,1) (P\_C,0.00699) (P\_D,0.753948) (P\_E,0.39829) (P\_F,0.044583)};
\addplot+[gray,fill=none, postaction={pattern=dots, pattern color=gray}] coordinates
{(P\_A,0.656772) (P\_B,0) (P\_C,0) (P\_D,0) (P\_E,0.60171) (P\_F,0.511435)};

\end{axis}

\node[rotate=90,anchor=center] at ([xshift=-3pt]current bounding box.west) {\small{\hspace*{0.5cm}Fraction of orig. prog. time}};
\end{tikzpicture}}
	\caption{Performance benefits due to \adapt{}}
	\label{fig:perf}
\end{figure}
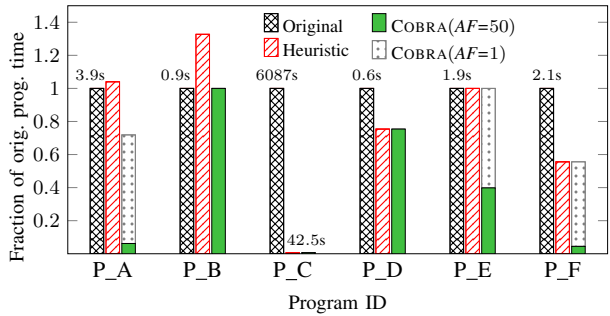

We ran \adapt{} on a representative sample from each category. 
\review{We used a data generator to generate test data based on the application schema, with the size of the largest relation(s) as 1 million. In particular, the following setup was used: fast local network, many to one mapping ratio 10:1, selectivity of any predicate used 20\%. Since we do not know the Wilos application characteristics to estimate the amortization factor, we evaluated \adapt{} with three different amortization factors (\textit{AF}=$1$, \textit{AF}=$50$, and \textit{AF}=$\infty$ ) in the cost model. The results for \textit{AF}=$50$ and \textit{AF}=$\infty$ were only marginally different, so for clarity, we only show the results for \textit{AF}=$1$ and \textit{AF}=$50$, in Figure~\ref{fig:perf}.}

The x-axis in Figure~\ref{fig:perf} shows the program identified
by its pattern ID, and the y-axis shows the fraction of the
actual execution time taken by a rewritten program 
in comparison to that of the original program. 
\review{We plot the following bars for each program. 
\textit{Original} \textendash{} the original program, 
\emph{Heuristic} \textendash{} program rewritten using 
the heuristic from~\cite{EMANI16} 
(push as much computation as possible into SQL query, 
then prefetch the query results at the earliest program point), 
\adapt{}$_{(\textit{AF=}50)}$ \textendash{} program rewritten using 
\adapt{} with AF=50, 
and \adapt{}$_{(\textit{AF=}1)}$ \textendash{} program rewritten 
using \adapt{} with AF=1.} The actual time in seconds for \emph{Original}
is shown above the bar. We use transformation rules proposed 
by earlier techniques~\cite{RAM12, EMANI16} (listed in Figure~\ref{fig:ir_trans}).

The results from Figure~\ref{fig:perf} suggest that performance benefits due to \adapt{} are significant. In the examples considered for this experiment, programs rewritten using \adapt{} gave up to 95\% improvement over the heuristic optimized program, when the cost was computed using \textit{AF}=$50$. Even with \textit{AF}=$1$, \adapt{} outperforms the original and heuristic optimized programs in some cases like A, as \adapt{}'s calculated iterative query invocations to be costlier and chose the prefetch alternative (refer Figure~\ref{fig:cases} pattern A). In cases B, C, and D, \adapt{} chose the same plan with \textit{AF}=$1$ as well as \textit{AF}=$50$, hence the bars are identical. Note that in each case, the program rewritten using \adapt{} (with \textit{AF}=$1$ or $50$) always performs at least as well as the original/heuristic optimized program. 
	
We now compare the plans (program implementations) chosen by the heuristic optimizer and \adapt{}. Remember that the heuristic optimizer pushes as much computation as possible into SQL. For programs that could entirely be translated into SQL (programs $C$ and $D$ in our workload), \adapt{} chose full SQL translation - same as the heuristic optimizer. For other programs where only a part of the program could be translated to SQL, \adapt{} differs from the heuristic optimizer. For instance, in program $A$ (nested loops with intermittent updates), the heuristic optimizer chose to translate the inner loop (which performs a filter) to SQL, whereas the outer loop could not be translated due to presence of updates. \adapt{} instead, chose to prefetch the inner loop query without the filter, thus eliminating iterative queries. Program $B$ contained two aggregations inside a loop on a query result - a scalar count, and a collection that accessed all the rows in the query result. While the heuristic optimizer translated the count computation into an additional SQL aggregate query, \adapt{} chose the original program with a single query. Programs $E$ and $F$ originally each contained SQL queries with a where clause, where the predicate differed. While this was deemed optimal by the heuristic optimzer, \adapt{} rewrote the queries without the where clause (similar to program A) to leverage multiple accesses to the same relation and employed prefetching.\\[3pt]

\noindent
\textbf{\adapt{} Optimization Time}:\\
The time taken for program optimization using \adapt{} is usually not a concern, as the program is optimized once for a particular environment and run multiple times. However, we note that in our experiments, the time taken for optimization was very small (\textless{}1s) for all programs.\\

\noindent
\review{\textbf{\emph{\underline{Threats to validity:}}} Our evaluation uses programs that use the Hibernate ORM as part of the Spring framework~\cite{SpringFrame}. Spring automatically takes care of transaction semantics based on annotations that specify which
functions are to be executed within a transaction. Each sample that we considered
in our evaluation runs under a single transaction (as is typical of a \textit{service} function in Spring), so cache invalidation across transactions
is not a problem. Further, Hibernate contains built in cache management for database mapped objects. In general for other database application
programs, optimizing across transactions may not preserve the program semantics and/or affect the amortization factor due to stale caches.
Identifying such cases automatically using program analysis is part of future work.}

\review{The values of parameters in our cost model have been tuned with respect to the Wilos application, which we used in our evaluation. However, in some cases, there was some deviation of the estimated program execution cost from the actual cost. We observed that this is due to multiple factors including (a) parameters not considered in the cost model (example: Hibernate's cost of constructing mapped objects from the result set), (b) fluctuating values of parameters (example: the utilized bandwidth is a fraction of the maximum bandwidth and varies across different query results), etc. Although our cost model correctly predicted the least cost alternative in all the evaluated samples despite these limitations, a more refined cost model may be desired in general.}

\section{Conclusion}
\label{sec:conclusion}
In this paper, we proposed a framework for generating 
various alternatives of a program using program transformations, 
and choosing the least cost alternative in a cost based manner. 
We identify that program regions provide a natural abstraction 
for optimization of imperative programs, and extend the 
Volcano/Cascades framework for optimizing algebraic expressions, 
to optimize programs with regions. Our experiments show that 
techniques in this paper are widely applicable in real world 
applications with embedded data access, and 
provide significant performance improvements.

Apart from various extensions identified throughout the paper, 
future work includes expanding the set of program transformations 
available in \adapt{}. 
For instance, program partitions in 
Pyxis~\cite{Cheung2012} can be modeled as partitions of regions in \adapt{}, 
with location as a physical property and enforcers to transfer data
between locations (these concepts already exist in Vol\-cano/Cas\-cades).
Although we have focused on data access optimizations for imperative programs, 
\adapt{} could be used for other cost based program transformations with
an appropriate cost model, an example being rewriting applications
for evolving schemas (Section~\ref{sec:relwork}). 

Another closely related problem is that of optimizing stored
procedures (SPs)/user defined functions(UDFs) in databases, which contain imperative 
constructs along with queries. Transformations for the same have been
proposed in earlier work~\cite{Simhadri14, Froid}. \adapt{} can be
used with these transformations, or independently, for cost based
optimization of SPs and UDFs.

\begin{figure*}
	\small
	\centering
	\begin{tabular}{|c|c|p{6cm}|}
		\hline
		\textbf{Sl.No.} & \textbf{Pattern ID} & \textbf{File Name (Line Number)}\\\hline
		1 & A & ProjectService (1139)\\
		2 & & TaskDescriptorService (198)\\
		3 & & ConcreteWorkBreakdownElementService (144)\\\hline
		4 & B & IterationService (139)\\
		5 & & PhaseService (185)\\\hline
		6 & C & ConcreteRoleAffectationService (60)\\
		7 & & ConcreteTaskDescriptorService (312)\\
		8 & & ConcreteTaskDescriptorService (1276)\\
		9 & & ConcreteTaskDescriptorService (1302)\\
		10 & & ConcreteWorkBreakdownElementService (63)\\
		11 & & ConcreteWorkProductDescriptorService (445)\\
		12 & & ParticipantService (129)\\
		13 & & RoleService (15)\\
		14 & & ActivityService (407)\\\hline
		15 & D & IterationService (293)\\
		16 & & PhaseService (307)\\
		17 & & ActivityService (229)\\
		18 & & RoleDescriptorService (276)\\
		19 & & TaskDescriptorService (140)\\
		20 & & TaskDescriptorService (142)\\
		21 & & WorkProductDescriptorService (310)\\\hline
		22 & E & ProjectService (346)\\
		23 & & ProjectService (567)\\
		24 & & ProjectService (647)\\
		25 & & ProjectService(704)\\
		26 & & ProcessService (1212)\\
		27 & & ProcessService (1253)\\
		28 & & ProcessService (1593)\\
		29 & & ProcessService (1631)\\
		30 & & ProcessService (1740)\\\hline
		31 & F &  ProcessService (406)\\
		32 &  & ProcessService (921)\\\hline
	\end{tabular}
	\caption{Code fragments for cost based rewriting}
	\label{fig:cases_details}
\end{figure*}

\bibliographystyle{IEEEtran}
\bibliography{references}

\appendix
\subsection*{Details of Code Fragments for Cost Based Rewriting}
\label{sec:cases_details}
We present the details of code fragments from Wilos~\cite{Wilos}
where cost based program transformations are applicable.
For each category identified in Figure~\ref{fig:cases}, 
Figure~\ref{fig:cases_details} lists all the code
fragments that contain the pattern of that category, along
with the file name and line number where the code fragment
occurs in the source code.
\end{document}